\documentclass[traditabstract]{aa}
\usepackage{graphicx}
\usepackage{txfonts}
\usepackage{courier}
\usepackage{pstricks}
\usepackage{natbib}
\usepackage{multirow}
\usepackage{color, colortbl}
\usepackage{hyperref}
\bibpunct{(}{)}{;}{a}{}{,}

\def\0{\phantom0}

\begin{document}

\title{A simple fitting method (\emph{gfit}) for galaxy shape\\  measurement in weak lensing surveys}

%   \subtitle{}

\titlerunning{}

\author{M.~Gentile\inst{1}, F.~Courbin\inst{1} \and G.~Meylan\inst{1}}

   \institute{Laboratoire d'astrophysique, Ecole Polytechnique
     F\'ed\'erale de Lausanne (EPFL), Observatoire de Sauverny,
     CH-1290 Versoix, Switzerland
  } 

\date{Submitted to \aap}
 
\abstract{
It is anticipated that the large sky areas covered by planned wide-field weak lensing surveys will reduce statistical errors to such an extent that systematic errors will instead become the dominant source of uncertainty. It is therefore crucial to devise numerical methods to measure galaxy shapes with the least possible systematic errors. We present a simple "forward deconvolution" method, \emph{gfit}, to measure galaxy shapes given telescope and atmospheric smearings, in the presence of noise. The method consists in fitting a single 2D elliptical S\'ersic profile to the data, convolved with the point spread function. We applied \emph{gfit} to the data proposed in the GRavitational lEnsing Accuracy Testing 2010 (GREAT10) Galaxy Challenge. In spite of its simplicity, \emph{gfit} obtained the lowest additive bias \mbox{($\sqrt{\mathcal{A}}=0.057\times10^{-4}$)} on the shear power spectrum
among twelve different methods and the second lowest multiplicative bias \mbox{($\mathcal{M}/2=0.583\times10^{-2}$)}. It remains that \emph{gfit} is a fitting method and is therefore affected by noise bias. However, the simplicity of the underlying galaxy model combined with the use of an efficient customized minimization algorithm allow very competitive performances, at least on the GREAT10 data, for a relatively low computing time.

%\abstract{
%It is anticipated that the large sky areas covered by planned wide-field weak lensing surveys will reduce statistical errors to such an extent that systematic errors will instead become the dominant source of uncertainty. It is therefore crucial to devise numerical methods to measure galaxy shapes with the least possible systematic errors. We present a simple "forward deconvolution" method, \emph{gfit}, to measure galaxy shapes given telescope and atmospheric smearings, in the presence of noise. The method consists in fitting a single 2D elliptical S\'ersic profile to the data, convolved with the point spread function. We applied \emph{gfit} to the data proposed in the GRavitational lEnsing Accuracy Testing 2010 (GREAT10) Galaxy Challenge. In spite of its simplicity, \emph{gfit} obtained the lowest additive bias \mbox{($\sqrt{\mathcal{A}}=0.057\times10^{-4}$)} on the shear power spectrum
%among twelve different methods and the second lowest multiplicative bias \mbox{($\mathcal{M}/2=0.583\times10^{-2}$)}. It remains that \emph{gfit} is a fitting method and is therefore affected by noise bias. However, the simplicity of the underlying galaxy model combined with the use of an efficient customized minimization algorithm allow very competitive performances, at least on the GREAT10 data, for a relatively low computing time.

} 

%\titlerunning{}
%%% Titre pour version referee
\maketitle
%
%________________________________________________________________
\section{Introduction}
\label{section:introduction}

 Weak gravitational lensing \citep[e.g.,][]{BartelmannSchneider2001, HoekstraJain2008}, whereby the gravitational bending of light by structures in the Universe slightly distorts images of distant galaxies,
 is now recognized as a powerful means to study the history of the Universe and probe the mysterious nature of the dark matter and dark energy \citep{Munshi2008, Huterer2010}. 
 %The lensing effect is very subtle, however, and requires measuring the shapes of thousands of faint galaxies. 
 %Moreover, before they reach the observer, the apparent galaxy images undergo a number of additional distortions, mainly due to the telescope optics, the light detector and the presence of noise.

Since the first detection of weak lensing \citep{Maoli2001, Bacon2000, Kaiser2000, vanWaerbeke2000, Wittman2000}, a number of methods have been devised and implemented to tackle the inverse problem of recovering the lensing signature from observed, distorted galaxy images \citep{KSB1995a, LuppinoKaiser1997, Hoekstra1998, BernsteinJarvis2002, HirataSeljak2003, RefregierBacon2003, HeymansSTEPI2006, MasseySTEP22007, Miller2007, Kitching2008, BridleGREAT082010, G10results}.

We describe in this paper \emph{gfit}, a simple shear measurement method that nevertheless obtained good results in the latest GRavitational lEnsing Accuracy Testing 2010 (GREAT10) Galaxy challenge \citep{GREAT10Handbook2010, G10results}. Galaxies are assumed to be well modeled by a seven-parameter, single-component elliptical S\'ersic profile. The shape measurement algorithm essentially consist in iteratively shearing and convolving the galaxy model until a sufficiently close match with the observed galaxy is reached. Instead of an out-the-box minimizer, we employ a custom-developed minimizer well suited to fitting faint and noisy images like those frequently found in weak lensing. 

The paper is structured as follows. We provide in Sect.~\ref{section:gfit method description} a description of the underlying principles, galaxy model and shape measurement \mbox{algorithm} of \emph{gfit}. We continue in Sect.~\ref{section:applying gift on g10 data} with a presentation of the pipeline we used to participate in the GREAT10 Galaxy challenge and follow with a analysis of the \emph{gfit} results in Sect.~\ref{section:result analysis}. We conclude in Sect.~\ref{section:conclusions}.

\section{The gfit shear measurement method}
\label{section:gfit method description}

\subsection{The shear measurement problem}
\label{subsection:shear measurement problem}

According to the theory of weak gravitational lensing, the light emitted by a galaxy is slightly deviated by the foreground gravitational field, an effect that can be modeled to first order as the combination of two effects, the \emph{convergence} $\kappa$ and the \emph{shear} $\gamma$, that describe how light bundles emitted by a source are distorted by a potential well.
The convergence models the magnification effect whereby the galaxy image see its apparent size increased without altering its shape, whereas the shear describes a stretching effect where only ellipticity is altered, not size.

All so-called ``shear measurement methods'' attempt to reconstruct the reduced shear $g=\gamma / ({1 - \kappa - \gamma})$ which is approximately equals to the shear $\gamma$ in the weak gravitational limit, where $\kappa \ll 1$ and $\gamma \ll 1$. 

The lensing effect is very subtle, however, and requires measuring the shapes of thousands of faint galaxies. Moreover, before they reach the observer, the apparent galaxy images undergo a number of additional distortions, unrelated to lensing, that further complicates that task, mainly:
%mainly due to the telescope optics, the light detector and the presence of noise.
%In theory, that shear could be directly estimated by measuring the deviation from circularity of a large number of galaxy shapes. But in practice a
%But, as illustrated in Fig. xxx, the task of extracting the shear is further complicated by a number of factors:
\begin{itemize}
%\item The convolution of the images by the instrumental and/or atmospheric point spread function (PSF), that alters the ellipticities due to the shear and flatten the galaxy light profile
\item The convolution of the images by the instrumental and/or atmospheric point spread function (PSF) that flattens and circularizes the galaxy light profile.
\item The Gaussian and Poisson noise introduced by the surrounding sky emissions and the detecting device.
\item The pixelation effect caused by the integration of light falling  on the detector pixels.
%\item The spatial variability of the shear field itself
\end{itemize}

The traditional approach for estimating the shear is to measure the deviation from circularity of a large number of galaxy shapes. But this technique assumes the shear remains constant across the field of view, which is generally not the case. Moreover, accurate shear measurement must also account for a spatially varying PSF that must be interpolated at the positions of the galaxies. 

A shear measurement pipeline must overcome all the above difficulties, typically going through the following steps:
\begin{enumerate}
\item \emph{PSF correction}, whose goal is to restore the shape a galaxy had before being convolved with the PSF. If the spatial variation of the PSF over the field of view is significant, this step also requires interpolating the PSF to the position of the galaxies in the sky.\\
\item \emph{Shape measurement}, that is, the estimation of the galaxy shape \emph{after} it has been altered by the cosmic shear but \emph{before} PSF convolution and other subsequent distortions. 
In this paper, we call the corresponding shape the \emph{sheared} galaxy shape. The real galaxy ellipticity prior to gravitational lensing (i.e. unsheared) is referred to as the \emph{intrinsic  \mbox{ellipticity}}.\\
\item \emph{Shear measurement}, that is, the task of extracting the shear signal from the sheared galaxy shapes estimated in the previous step. A spatially-varying shear field is commonly described as a power spectrum or a correlation function. 
\end{enumerate}
Additional steps may also be performed to correct the images from the effects of pixelation and noise.

\subsection{Shear measurement with gfit}
The \emph{gfit} ``galaxy fitting'' method grew from a prototype initially developed by St\'ephane Paulin-Henrikson on the occasion of the GREAT08 challenge \citep{BridleGREAT08Handbook2008}, where  it obtained the third and fifth ranks on images with high and low signal to noise ratios respectively \citep{BridleGREAT082010}. 

The GREAT10 \emph{gfit} code was subsequently made more generic in order to satisfy the more demanding requirements of the Galaxy challenge. It was also enhanced in several aspects that we describe in subsequent sections of this paper.

%The \emph{gfit} ``galaxy fitting'' method was initially developed at the occasion of the GREAT08 challenge \citep{BridleGREAT08Handbook2008}, in collaboration with St\'ephane Paulin-Henrikson [CHECK] where it obtained respectively the third and fifth ranks on images with high and low signal to noise ratios \citep{BridleGREAT082010}. 
% MOVED to the GREAT10 part
%With \emph{gfit}, PSF correction and shape measurement, i.e. steps~(1) and (2) above, are performed simultaneously using a forward-fitting algorithm that we describe in Sect.~\ref{subsection:gfit algorithm}. 
%In GREAT10, step~(3), i,e., the estimation of the shear field from the sheared galaxy shapes, was not mandatory and the participants were allowed to supply for each image a catalog of sheared ellipticities instead of a shear power spectrum \citep{GREAT10Handbook2010}. Like most other competing methods, \emph{gfit} only provided its estimates in the form of a catalog of ellipticities at requested positions within the images. Future version of \emph{gfit} will allow the extraction of a spatially varying shear. 

It is also worth mentioning that a wavelet-based denoising algorithm described in \citet{Nurbaeva2011} was also experimented in the Galaxy challenge and proved quite successful. This denoising scheme is presented in Sect.~\ref{subsection:denoising}.\\

The overall shape measurement procedure is the following:
\begin{enumerate}
\item Application of the denoising algorithm on the galaxy and/or PSF images [optional]
\item Estimation of the galaxy and PSF centroids in all images
\item Application of the PSF correction and shape measurement algorithm
\item Generation of the ellipticity catalogs
\item Production of various statistics and plots for analysis [optional]
\end{enumerate}

We detail in the next sections the shape measurement algorithm along with its underlying models and components.

\subsection{Modeling the galaxies}
\label{subsection:galaxy model}

\emph{gfit} is fundamentally a model-fitting method. We describe here the model used to represent galaxies and cover the fitting-related aspects in Sects.~\ref{subsection:gfit algorithm} and \ref{subsection:minimizer}.\\

Galaxies are assumed to have a surface brightness distribution well described by an elliptical S\'ersic function  \citep{Sersic1968}, defined by:

\begin{equation}
\label{eq:eliptical Sersic function} I(\xi,n,re)=I_{sky} + I_0\,\exp{\big[-b_n\,\big({\frac{\xi}{r_e}}\big)^{1/n} \big]} 
\end{equation} where:

\begin{itemize}
\item $I_{sky}$ represents the sky brightness
\item $I_0$ is the central surface brightness of the galaxy
\item $n$ denotes the S\'ersic index that determines the degree of curvature of the profile. A small value of $n$ leads to a less centrally concentrated profile and a shallower logarithmic slope at small radii. 
\item The scale radius parameter $r_e$ is defined as the effective radius encircling half of the total light of the profile \citep[e.g.][]{Ciotti1991, TrujilloErwinRamosGraham2004, GrahamDriver2005}. 
\item  The factor $b_n$ \citep[e.g.][]{CiottiBertin1999} arises from the definition of $r_e$ and is related to the S\'ersic index $n$ through the equation $\Gamma(2\,n) = 2\,\gamma\,(2n,b_n)$ where $\Gamma$ and $\gamma$ functions are respectively the complete and incomplete gamma functions \citep{AbramowitzStegun1965}. 
\item The parameter $\xi$ in Eq.~(\ref{eq:eliptical Sersic function}) is defined as \begin{equation}
\xi=\sqrt{(x'-x_c)^2+\frac{(y'-y_c)^2}{q^2}}
\label{eq:radius distance} 
\end{equation} and denotes the distance from the centroid $(x_c,y_c)$ of the galaxy to a point on an elliptical isophote at spatial coordinate $(x', y')$
\begin{displaymath}
 \left[ \begin{array}{c} x'-x_c\\ y'-y_c\\ \end{array} \right] = 
 \left[ \begin{array}{rl} \cos{\phi} & \sin{\phi} \\ -\sin{\phi} & \cos{\phi} \\  \end{array} \right]\,
 \left[ \begin{array}{c} x-x_c \\ y-y_c \\ \end{array} \right],
\end{displaymath}
obtained after \mbox{counterclockwise} rotation through an angle $\phi$ with respect to the $(0,\,x)$ axis.
\item The quantity $q$ in expression~(\ref{eq:radius distance}) of $\xi$ is the ratio of the semi-minor axis $b$ to the semi-major axis $a$ of the isophote ellipse. It is related to the complex ellipticity $\mathbf{e}=(e_1,e_2)$ of the galaxy through $q=b/a=(1-\arrowvert\mathbf{e}\arrowvert)/(1+\arrowvert\mathbf{e}\arrowvert)$ with $\arrowvert\mathbf{e}\arrowvert=\sqrt{{e_1}^2+{e_2}^2}$, $e_1=\arrowvert\mathbf{e}\arrowvert\,\cos 2\phi$ and $\,e_2=\arrowvert\mathbf{e}\arrowvert\,\sin 2\phi$. 
\end{itemize}
This model was initially chosen for its simplicity and has the other merit of being relatively easy to fit. A galaxy represented as the sum of a bulge and a disc would be more realistic, but it is not clear whether such a model would prove more accurate on weak lensing images and worth the additional complexity and computational cost, given degeneracies between parameters. As regards \emph{gfit}, the GREAT10 results does not provide a definitive answer (see Sect.~\ref{section:result analysis}).
%This model was initially chosen for its simplicity and has the other merit of being relatively easy to fit. A galaxy represented as the sum of a bulge and a disc would be more realistic, but it is not clear whether such a model would prove more accurate on weak lensing images. And if it were the case, the actual gain in accuracy would possibly not be worth the additional complexity and computational cost. As regards \emph{gfit}, the GREAT10 results does not provide a definitive answer, as shown in Sect.~\ref{subsection:analysis}.

\subsection{Galaxy shape measurement and PSF correction}
\label{subsection:gfit algorithm}

We describe in this section the shape measurement algorithm of \emph{gfit}. It is based on iterative fitting of observed galaxies to the galaxy model described in Sect.~\ref{subsection:galaxy model}.
The basic assumptions of the algorithm are the following:
\begin{itemize}
\item All galaxy and PSF fields have been reduced and the objects they contained assumed available in the form of square postage stamps in FITS format. We denote by $O$ a galaxy field image of the GREAT10 Challenge and by $o$ any of its 10,000 galaxy postage stamps. The GREAT10 PSF field image corresponding to $O$ is labeled as $P$ and any of its individual PSF kernel as $p$.
\item  The $O$ and $P$ images have been denoised beforehand if requested. The denoising tool described in Sect.~\ref{subsection:denoising} is the default choice.
\item The centroids of all galaxies and PSF to be processed have been estimated. A tool based on SExtractor \citep{SExtractor1996} has been written for this purpose (see Sect.~\ref{subsection:centroid estimation}.)
\item The PSF field at the position of each galaxy is known, so that we can always find the PSF $p$ that matches galaxy $o$.
\end{itemize}
%
%The algorithm takes as input
%\begin{itemize}
%\item A FITS  file containing a set of observed galaxy in the form of squared postage stamps. The mages are assumed  noisy, pixelized and convolved with a known PSF.
%\item A FITS file  The PSF best matching the galaxy, also in the form of a postage stamp.
%\item 
%\end{itemize}
The fitting algorithm itself is summarized below:

\renewcommand{\labelitemiii}{$\Rightarrow$}
\renewcommand{\labelitemii}{$\diamond$}
\begin{itemize}

\item For each galaxy object $o$ in observed galaxy field $O$
\begin{itemize}
\item Extract a square stamp cutout of a given dimension $N_G \times N_G$ around the centroid $(x_c,y_c)$ of $o$.
\item Remove the sky background from $o$ after having estimated its variance ${\sigma^2_{sky}}$.
\item Estimate the galaxy noise variance ${\sigma}^2$ of $o$.

\item Select guess parameters for the 7 model parameters $\big\{I_0, (x_c,y_c), (e_1,e_2),  n, r_e$\big\} described in Sect.~\ref{subsection:galaxy model}.
\item Construct an initial galaxy postage stamp $g$ based on the S\'ersic profile having these parameters.
\item Select the PSF $p$ that matches the spatial coordinates of $o$ in $O$ and extract a cutout of dimension $N_P \times N_P$ about the centroid. Make sure the PSF is normalized and has the sky background removed.
\item Iteratively vary the model parameters with the objective of minimizing the residuals between $g$ and $o$. \\ 
At each step:

\begin{itemize}
\item Convolve the model galaxy $g$ with the observed PSF $p$ and compute a new galaxy estimate $k~=~p \star g$
\item Compute the residuals between $o$ and $k$ using the chi-squared statistics
\begin{displaymath}
%{\chi^2=\sum\limits_{i}\frac { {(o - k)}^2} {\sigma_i}^2}  }
{\chi^2=\sum\limits_{n}\sum\limits_{m}\frac{{\big(o_{n,m} - k_{n,m}\big)}^2} {{\sigma^2_{n,m}}}}
\end{displaymath} where $o_{n,m}$ and $p_{n,m}$ represent the pixel value at position $(n,m)$ in $o$ and $p$ respectively. Similarly, $\sigma_{n,m}$ denote the standard deviation of the noise associated with each pixel at $(n,m)$.
%with ${\sigma_i}=\sqrt{{{(\sigma_n}^2 + \sigma_b)}^2}$
%$ {\chi^2=\sum\limits_{i}\frac{{[o - k]}^2} {{\sigma_i}^2}}$ with ${\sigma_i}=\sqrt{{{(\sigma_n}^2 + \sigma_b)}^2}$ \\ 
\item If the minimum $\chi^2$ has not been reached, select a new set of parameters and construct the corresponding model galaxy $g$. Otherwise, exit the minimization loop.
\end{itemize}
%\item To perform the minimization steps above, \emph{gfit} uses its own minimizer described in Sect.~\ref{subsection:minimizer}.  \end{itemize}
\item At the end of the iteration cycle, the algorithm yields: 

\begin{itemize}
\item An estimate for all 7 model parameters and for the complex ellipticity $(e_1,e_2)$ in particular. 
%Quantities such as the ellipticity modulus, the position angle or the axis ratio can be easily derived.
%The ellipticity modulus can be derived using $\arrowvert\mathbf{e}\arrowvert=\sqrt{{e_1}^2+{e_2}^2}$, the position angle as $\phi=\frac{1}{2}\arctan{(e_2 / e_1)}$ and the axis ratio as $q=({1-\arrowvert\mathbf{e})/(\arrowvert}/{1+\arrowvert\mathbf{e}\arrowvert})$.
\item A postage stamp for the best-fitted intrinsic galaxy postage stamp $g$.
\item A postage stamp for the best-fitted convolved galaxy $k$.
\end{itemize}
\item The same procedure is followed for the next galaxy $o$ in $O$, until all galaxies have been processed. \\ 
\end{itemize}
\item Once the shape of all galaxies have been measured, produce a catalog containing the fitted parameters at the positions of the galaxies $o$ in $O$ and optionally, additional statistics and plots. Quantities such as the ellipticity modulus $\arrowvert\mathbf{e}\arrowvert$, the position angle $\phi$ or the minor-to-major axis ratio $q$ can be respectively derived from $(e_1,e_2)$ using $\arrowvert\mathbf{e}\arrowvert=\sqrt{{e_1}^2+{e_2}^2}$, $\phi=\frac{1}{2}\arctan{(e_2 / e_1)}$ and $q=({1-\arrowvert\mathbf{e}\arrowvert})/({1+\arrowvert\mathbf{e}\arrowvert})$.
\end{itemize}

The algorithm just described is not dependent on the galaxy or the PSF model. It would remain unchanged, for instance, if the centroid of the PSF itself was taken into account during fitting or if a bulge and a disc were incorporated in the galaxy model. The quality of the whole shape measurement procedure can thus be increased by improving the models.
%It is worth noting that:
%\begin{itemize}
%\item The algorithm just described is not dependent on the galaxy or the PSF model. I would remain unchanged, for instance, if the centroid of the PSF itself was taken into account during fitting or if a bulge and a disc were incorporated in the galaxy model. The quality of the whole shape measurement procedure can thus be improved by improving the models and/or improving the algorithm itself.
%\item By default the 
%\end{itemize}

The shape measurement algorithm has its own strengths and weaknesses, outlined below: 
\begin{itemize}
\item Strengths:
\begin{itemize}
\item Simplicity.
\item PSF deconvolution and galaxy model estimation are performed simultaneously.
\item The deconvolution process does not involve any matrix inversion and its side effects (numerical instabilities, presence of artifacts, noise amplification, etc.).
\item The intrinsic and best-fitted modeled galaxies are obtained as a by-product of the algorithm, in addition to the estimated model parameters.
\end{itemize}
\item Weaknesses:
\begin{itemize}
\item A bias in introduced through the choice of a galaxy model, which is necessarily imperfect.
\item The choice of the initial guess parameters of the galaxy model are more or less arbitrary and may also influence the final model parameter estimates:
\item The algorithm relies heavily on the accuracy and robustness of the minimizer.
\item The methods is sensitive to noise bias
\item No estimate of errors is available
\end{itemize}
\end{itemize}

\subsection{The gfit minimizer}
\label{subsection:minimizer}

The \emph{gfit} galaxy model expressed by Eq.~(\ref{eq:eliptical Sersic function}) varies linearly with the parameter $I_0$, but non-linearly with the remaining parameters $\big\{(x_c,y_c), (e_1,e_2), n, r_e\big\}$. This requires the use of a non-linear minimization algorithm over a seven-dimensional parameter space. 

Unlike linear minimizations schemes that only involve a matrix inversion, non-linear optimization requires iterating over the parameter space to find the minimum value of the objective function, which is in our case the $\chi^2$ of the residuals between observed and estimated images. That minimum is not necessarily the absolute minimum of the $\chi^2$ function but the most relevant from the point of view of the physics of the problem. In our case, the minimum should coincide with the S\'ersic model parameters that best fit the galaxy shape.

%The \emph{gfit} galaxy model expressed by Eq.~(\ref{eq:eliptical Sersic function}) varies linearly with the parameter $I_0$, but non-linearly with the remaining parameters $\big\{(x_c,y_c), (e_1,e_2), n, r_e\big\}$. This requires the use of a non-linear minimization algorithm over a seven-dimensional parameter space. Unlike linear minimizations schemes that only involve a matrix inversion, non-linear optimization requires iterating over the parameter space to find an appropriate $\chi^2$ minimum. That minimum is not necessarily the absolute minimum of the $\chi^2$ function but the most relevant from the point of view of the physics of the problem. In our case, the minimum should coincide with the S\'ersic model parameters that best fit the galaxy shape.

A good minimizer is essential to any galaxy model fitting algorithm, but finding such a minimum in a reliable manner can prove tricky for a number of reasons:
\begin{itemize}
\item Parameter degeneracy, where different combinations of the parameters yield similar $\chi^2$ values.
%\item Coupling between parameters in the model, so that all parameters are not independent and must be fitted together.
%\item Some parameters may have a greater influence on the objective function than others.
%\item Coupling between parameters in the model, where one he value of one parameter may in turn depend on those of other parameters.
%\item Effect of noise, undersampling and measurement errors that can degrade the accuracy of the fit or may even alter the galaxy shape so much that convergence fails. 
\item Errors related to noise and undersampling may degrade the accuracy of the fit in several ways: (1) the distortion of the galaxy image may be such that the minimizer fits a wrong shape, even if it does it accurately. (2) The minimizer may be fooled by a ``false'' local minimum and converge toward wrong fitted parameter values. (3) The minimizer may fail to converge altogether if it cannot reconcile the model with the observed image.
\item The choice of initial guess values for the parameters can influence the outcome of the minimization algorithm: depending on the starting location on the parameter hypersurface, the algorithm may tend to follow a different path and converge to a different minimum.
\end{itemize}

In this regards, the choice of the galaxy model used by \emph{gfit} leads to a number of challenges:
\begin{itemize}
\item The total flux $I_0$ is degenerate with the $n$ and $r_e$ parameters, so that, for instance, in Eq.~(\ref{eq:eliptical Sersic function}), a low $I_0$ flux may be compensated by a higher exponential function of $n$, $re$ and $(x_c,y_c)$, modifying the shape of the fitted galaxy profile. Similarly, an error in the estimation of the sky brightness $I_{sky}$  may ``drive'' the minimizer toward a wrong combination of the remaining parameters. This is why we estimate $I_{sky}$ separately in the algorithm described in Sect.~\ref{subsection:gfit algorithm}.
%\item The parameters $I_0$, $n$ and $re$ are strongly coupled. 
\item \emph{gfit} attempts to limit the effect of noise by applying the denoising method described in Sect.~\ref{subsection:denoising}. As regards pixelization, \emph{gfit} can optionally construct the galaxy model in higher resolution and rebin the pixels before fitting.
%\item The total flux $I_0$ and the sky brightness $I_{sky}$ are degenerate with the remaining parameters the $n$ and $r_e$ parameters. For instance a low flux may be compensated by a larger exponent $n$
\item To estimate the initial guess values for fitting a particular galaxy, \emph{gfit} can be setup to either accept default values or use estimates from SExtractor.
\end{itemize}

Several families of optimization algorithms were experimented on GREAT08 and GREAT10 images: simplex (\emph{\mbox{Nelder-Mead downhill}}), gradient descent (\emph{Powell}), Newton \& quasi-Newton (\emph{Newton-CG}, \emph{BFGS}) and \emph{Levenberg-Marquardt} (LVM). Descriptions of these algorithms can be found in \citep[e.g.][]{Levenberg1944, Marquardt1963, Powell1964, NelderMead1965, ZhuByrdLuNocedal1997, NocedalWright1999,BonnansGilbertLemarechalSagastizabal2006}.

None of these methods proved entirely satisfactory, either failing to converge or yielding insufficient accuracy, especially on low signal to noise ratio (S/N) images.
The LVM implementation from the SciPy library \citep{Scipy, ScipyCommunity2010} that we used was the fastest and the most accurate. For these reasons we used it in the GREAT08 version of \emph{gfit}, but that implementation of LVM:
\begin{itemize}
\item Failed to converge in about 5-10\% of the time on GREAT08 ``real noise blind''  images.
\item Was occasionally tricked by ``false'' local minima, producing the smallest residuals but with unphysical S\'ersic parameters or ellipticities.
\item Required good estimates of guess parameters in order to converge towards the right minimum.
\end{itemize}
We also experimented a parameter estimation scheme based on a Bayesian approach and implemented using the \emph{pymc} Markov Chain Monte Carlo (MCMC) library of \citet{PatilHuardFonnesbeck2010}, but that method produced less accurate estimates while being less computationally efficient.

In an attempt to overcome these issues, we eventually decided to implement a custom minimizer, better suited to fitting noisy, pixelized galaxy images than vanilla minimization algorithms. We found that a scheme based on an adaptive \emph{cyclic \mbox{coordinate descent}} algorithm (CCD) was able to produce more accurate estimates while at the same time being more robust:
\begin{itemize}
\item Able to better cope with degenerate and correlated parameters: LVM has difficulties with the $\big\{I_0, n, r_e\big\}$ degeneracy and coupling and the steepest descent algorithm used in LVM occasionally jumps without precaution to a minimum value that may be the smallest but not the most appropriate one. To avoid this, the CCD algorithm ensures that these parameters are carefully varies in a ``round-robin'' manner at the beginning of the fitting process, where the amplitudes in variation are the greatest. This scheme is also much more tolerant with regards to initial guess values and converges reliably, making it more robust overall.
\item More resilient to noise and on average more accurate than LVM, especially on low S/N images.
\end{itemize}
The CCD algorithm proved suitable for fitting without any single failure the huge number of GREAT10 galaxies. It nevertheless has a number of drawbacks, namely:
\begin{itemize}
\item The convergence rate is lower than that of LVM, resulting in a greater number of function evaluation. It is thus much slower than LVM.
\item Its efficiency decreases rapidly with the number of parameters. The algorithm performance is also influenced by factors such as the specific stopping conditions chosen or the range if iteration step sizes specified for the parameters.   
 
\end{itemize}
%The main features of the gfit minimizer are summarized in Table~xxx.

The current gfit implementation can be configured to use either the CCD, LVM or MCMC-based minimizer.

\subsection{Centroid estimation} 
\label{subsection:centroid estimation}

\emph{gfit} does not assume objects to be correctly centered within their postage stamps and accurate estimates for galaxy and PSF objects are required for two main reasons: 
\begin{itemize}
\item \emph{gfit} does not necessarily use the whole postage stamps for the galaxy and the PSF to save computational time and reduce noise: the corresponding postages stamps are cut out to a smaller dimension (e.g. $24 \times 24$ instead of e.g., $48 \times 48$ around the estimated centroid.
\item The coordinates of the galaxy and PSF centroids are provided to the minimizer as initial guess values before the fitting cycle can begin. Accurate centroids are especially important if the LVM minimization algorithm is used (CCD is much more tolerant in this respect).
\end{itemize}
\emph{gfit} relies on centroid estimates obtained from the SExtractor tool \citep{SExtractor1996}. A catalog is generated with centroid information and additional data such as flux, ellipticities and position angles, that can be optionally used to set guess parameter values. 
%A tool has been written to process multiple images in parallel with SEXtractor, either on machines with symmetric multiprocessing (SMP) architecture or on parallel computers via the Message Passing Interface (MPI) interface \citep{MPI11995, MPI21998}.

\subsection{Denoising}
\label{subsection:denoising}

Correcting astronomical images from the effect of noise has always been a challenging task. This is particularly true for galaxy images captured for weak lensing analysis because noise not only degrades the overall quality of these images but also alters the shapes of the galaxies. This causes serious difficulties to all existing shear measurement schemes that found their shear extraction algorithms on the accurate measurement of galaxy shapes. 

The challenge is then to correct galaxy images from noise without compromising the shear signal they encode. 
Unfortunately, popular denoising algorithms based on median filtering \citep{Arce2005, AriasDonoho2009}, Wiener filtering \citep{Wiener1949, Khireddine2007} or discrete wavelet transform (DWT) \citep{BruceDonohoGaoMartin1994, Vetterli1995} are not shape preserving and do not meet that requirement.

%The laboratory of astrophysics of EPFL has recently developed \emph{DWT-Wiener}, a shape-preserving denoising technique combining DWT and Wiener filtering \citep{Nurbaeva2011}. That algorithm has been experimented during the GREAT10 Galaxy challenge and improved the quality factors of all the shear measurement methods that participated in the Galaxy challenge \citep{G10results}: that was the case for \emph{gfit} but also for \emph{MegaLUT} \citep{Teves2012} and \emph{TVNN} \citep{Nurbaeva2012}. 
By default, \emph{gfit} uses \emph{DWT-Wiener}, a shape-preserving denoising technique combining DWT and Wiener filtering developed at the laboratory of astrophysics of EPFL by \citet{Nurbaeva2011}. That algorithm has been experimented during the GREAT10 Galaxy challenge and was able to significantly improve the quality factors of all the shear measurement methods from EPFL that participated in the Galaxy challenge \citep{G10results}: that was the case for \emph{gfit} but also for \emph{MegaLUT} \citep{Tewes2012} and \emph{TVNN} \citep{Nurbaeva2012}. 

Interestingly, denoising improves the shape measurement of the three algorithms even though they are fundamentally different from each other.

Beyond shape-preservation, another advantage of the \emph{\mbox{DWT-Wiener}} algorithm lies in its ability to denoise ``in one go'' images containing a great number of objects, without having to individually process each object in turn. In GREAT10, for instance, \emph{\mbox{DWT-Wiener}} was directly applied to images containing $100 \times 100$ PSF or galaxy postage stamps. 

\section{Applying gfit to the GREAT10 data}
\label{section:applying gift on g10 data}

\subsection{The GREAT10 Galaxy challenge}
\label{subsection:G10 galaxy challenge}

We describe in this section the specific pipeline we used in the GREAT10 Galaxy Challenge competition that took place between December 2010 and September 2011. The aim, content and rules of the challenge have been described in the GREAT10 Handbook \citep{GREAT10Handbook2010, G10results}. In a nutshell,  the main goals of the GREAT Challenges are (i) to test existing weak lensing measurements methods and (ii) to promote the development of new, more accurate, shear measurement techniques.\\

\begin{figure}[t]
\resizebox{\hsize}{!}{\includegraphics[trim=0mm 0mm 0mm 0mm, clip]{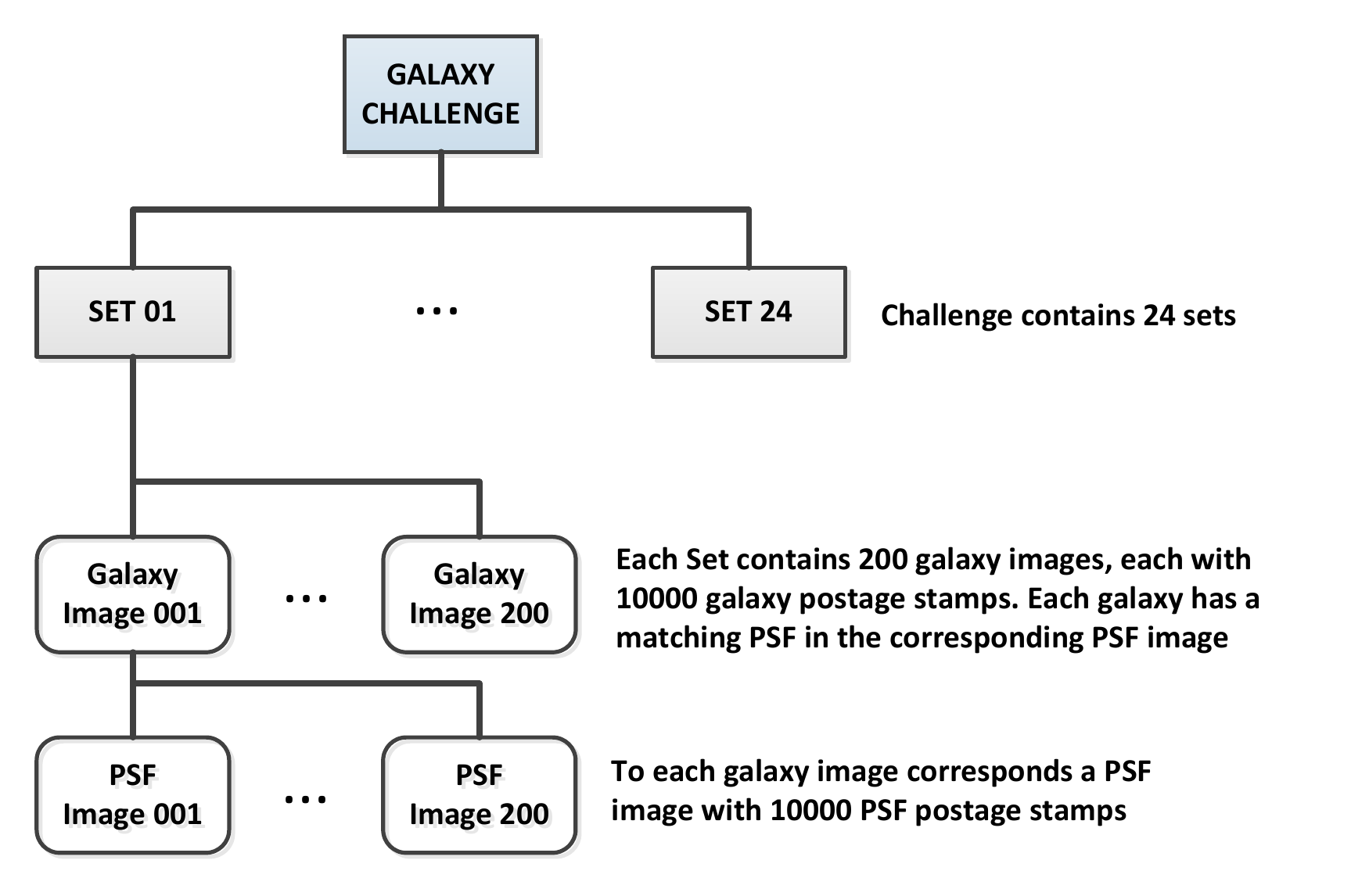}}
\caption{The GREAT10 Galaxy challenge dataset structure. There are 24 galaxy sets, each containing 200 galaxy images with 10000 galaxy postage stamps each. To each galaxy postage stamp corresponds a matching PSF postage stamp at the same spatial position in the galaxy image.}
\label{fig:dataset structure}
\end{figure}

The data consist of 24 datasets of 200 simulated galaxy images, each containing 10,000 noisy, PSF-convolved $48 \times 48$ pixel galaxy postage stamps, arranged on a $100 \times 100$ grid (see Fig.~\ref{fig:dataset structure}).  The GREAT10 edition includes spatially-varying PSF and shear fields, contrary to its predecessor, the GREAT08 challenge  \citep{BridleGREAT08Handbook2008, BridleGREAT082010}, where these fields were set as constant.

Each of the 24 sets is designed to evaluate the ability of competing methods to deal with galaxy or PSF fields with different properties (e.g. size, signal to noise ratio). We have reproduced in Table~\ref{table:property per set} the main PSF and galaxy characteristic attached to each of the GREAT10 set, as specified in the Galaxy Challenge results paper \citep{G10results}, Appendix D. 

The sets were also classified into ``Single epoch'', ``Multi-epoch'' and ``Stable single epoch'', depending on whether the intrinsic ellipticities and PSF keep the same or change their spatial distribution between images in a set (see Table~\ref{table:spatial variability}). 

Participating methods were ranked according to the following metrics :
\begin{itemize}
\item A ``Raw''  quality factor $Q$ according to which the live leader board was scored and that measures the difference, averaged over all sets, between the reconstructed and true shear power spectra.
\item A quality factor $Q_{dn}$ obtained after estimation and correction of pixel noise.
%\item A quality factor $Q_{dn}$ obtained after removal of the biases caused by finite signal-to-noise or inherent shape measurement method-related noise.
\item A quality factor $Q_{dn\, \&\, train}$ after application of an additional training step on top of pixel-denoising. That step consists essentially in estimating the multiplicative and additive biases on high S/N galaxy images (set 7) and applying that calibration on the remaining sets.

\item A split of the total bias into an shear-independent additive bias and a shear-dependent multiplicative bias between the measured and true shear, respectively denoted as $c$ and $m$ where $c=\langle c_i \rangle$ and $m=\langle m_i \rangle$, $i=(1,2)$. The $c_i$ and $m_i$ are respectively the additive and multiplicative biases over the two component of shear $\gamma_1$ and $\gamma_2$ and are similar to those used in STEP I \citep{HeymansSTEPI2006} and GREAT08 \citep{BridleGREAT082010} for a constant shear field. They are calculated by assuming the error on the shear, $\gamma_i - \gamma_i^{true}$, and the true shear $\gamma_i^{true}$ obeys a linear relationship of the form
\begin{equation}
\gamma_i - \gamma_i^{true} = m_i\,\gamma_i^{true} + c_i \qquad (i=1,2)
\end{equation} More details are provided in Appendix B of \citep{G10results}. See e.g. \citet{Huterer2006} for a discussion on the origin of additive and multiplicative errors in weak lensing studies. 

\item Additional bias metrics $\mathcal{A}\simeq\sigma^2(c)$ and $\mathcal{M} \simeq m^2 + 2m$, intended to measure the additive and multiplicative biases calculated at power spectrum level. Unlike $c$ and $m$, these metrics account for spatial variability.

\end{itemize}

\begin{table}[b]
\caption{Spatial variability of PSF and intrinsic ellipticities: images within a set may keep or not the same PSF or intrinsic ellipticity pattern of variation.}
\renewcommand{\arraystretch}{1.30}
\label{table:spatial variability}
%\begin{center}
\begin{tabular}{lll}
\hline
Type of variability within a set & PSF & Intrinsic ellipticity\\ 
\hline
Type 1: ``Single epoch'' & Variable & Variable \\
Type 2: ``Multi-epoch'' & Variable & Fixed \\
Type 3. ``Stable single epoch'' & Fixed & Variable \\

\hline
\end{tabular}
\end{table} 

\definecolor{FullVariability}{rgb}{0.824,0.824,0.824}

\begin{table}
\caption{Some of the PSF and galaxy properties characterizing the GREAT10 image sets. The second and third columns specify whether the PSF or intrinsic ellipticity field were kept constant for all images within a set. The parameters in the fourth column have been detailed in \citet{G10results}. The default signal to noise (S/N) ratio is 20, while low and high S/N ratios are 10 and 40 respectively.  All sets, except the last four, have galaxies with co-centered bulges and disks with a 50/50 bulge-to-disk ratio.}
\renewcommand{\arraystretch}{1.20}
\label{table:property per set}
%\begin{center}
\begin{tabular}{llll}
\hline
Set & PSF & Intrinsic ellipticity & Property of images \\ 
\hline
\cellcolor{FullVariability}1 & \cellcolor{FullVariability}Variable &\cellcolor{FullVariability}Variable &\cellcolor{FullVariability}Fiducial  \\
2 & Fixed & Variable & Fiducial  \\
3 & Variable & Fixed & Fiducial \\
\cellcolor{FullVariability}4 & \cellcolor{FullVariability}Variable &\cellcolor{FullVariability}Variable &\cellcolor{FullVariability}Low S/N  \\
5 & Fixed & Variable &  Low S/N \\
6 & Variable & Fixed &  Low S/N  \\
\cellcolor{FullVariability}7 & \cellcolor{FullVariability}Variable &\cellcolor{FullVariability}Variable &\cellcolor{FullVariability}High S/N  \\
8 & Fixed & Variable &  High S/N  \\
9 & Variable & Fixed & High S/N  \\
\cellcolor{FullVariability}10 & \cellcolor{FullVariability}Variable &\cellcolor{FullVariability}Variable &\cellcolor{FullVariability}Smooth S/N  \\
11 & Fixed & Variable &  Smooth S/N  \\
12 & Variable & Fixed & Smooth S/N  \\
\cellcolor{FullVariability}13 & \cellcolor{FullVariability}Variable &\cellcolor{FullVariability}Variable &\cellcolor{FullVariability}Small galaxy  \\
14 & Fixed & Variable & Small galaxy \\
\cellcolor{FullVariability}15 & \cellcolor{FullVariability}Variable &\cellcolor{FullVariability}Variable &\cellcolor{FullVariability}Large galaxy  \\
16 & Fixed & Variable &  Large galaxy \\
\cellcolor{FullVariability}17 & \cellcolor{FullVariability}Variable &\cellcolor{FullVariability}Variable &\cellcolor{FullVariability}Smooth galaxy  \\
18 & Fixed & Variable &  Smooth galaxy \\
\cellcolor{FullVariability}19 & \cellcolor{FullVariability}Variable &\cellcolor{FullVariability}Variable &\cellcolor{FullVariability}Kolmogorov PSF  \\
20 & Fixed & Variable & Kolmogorov PSF \\
\cellcolor{FullVariability}21 & \cellcolor{FullVariability}Variable &\cellcolor{FullVariability}Variable &\cellcolor{FullVariability}Uniform bulge/disc ratios \\
22 & Fixed & Variable &  Uniform bulge/disc ratios \\
\cellcolor{FullVariability}23 & \cellcolor{FullVariability}Variable &\cellcolor{FullVariability}Variable &\cellcolor{FullVariability}50/50 bulge/disc offset\\
24 & Fixed & Variable & 50/50 bulge/disc offset \\
\hline
\end{tabular}
%\end{center}

\end{table}

\subsection{The GREAT10 gfit implementation}
\label{subsection:gfit G10 implementation}

The GREAT10 version of \emph{gfit} only implements the first two steps described in Sect.~\ref{subsection:shear measurement problem}, that is, PSF correction and galaxy shape measurement.

In GREAT10, the estimation of the shear field (third step in Sect.~\ref{subsection:shear measurement problem}) was not mandatory as participants were allowed to supply for each image a catalog of estimated galaxy ellipticities instead of a shear power spectrum \citep{GREAT10Handbook2010}: an analysis program was written by the GREAT10 team to calculate a shear power spectrum from user-supplied ellipticity catalogs. Consequently, like most other competing methods, \emph{gfit} only provided its estimates in the form of a catalog of estimated ellipticities at requested positions within the images. Future version of \emph{gfit} will allow the extraction of a spatially varying shear. 

The \emph{gfit} implementation used in GREAT10 consisted of the following stages:
\begin{enumerate}
\item Optional denoising of the galaxy and PSF images with the DWT-Wiener method presented in Sect.~\ref{subsection:denoising}.

\item Centroid estimation using SExtractor \citep{SExtractor1996} as described in Sect.~\ref{subsection:centroid estimation}.

\item Galaxy shape measurement with the \emph{gfit} program, configured to use the CCD minimization algorithm described in Sect.~\ref{subsection:minimizer}. We used $24\times24$ pixel cutouts instead of the full $48\times48$ pixel original galaxy postage stamps.  Similarly, the size of PSF postage stamps was reduced to $12\times12$ pixels. That decision was made in order to keep the \mbox{overall} computation time within acceptable limits and avoid picking-up too much noise near the borders of the postage stamps.
\end{enumerate}

Because of the large number of galaxies, running the pipeline on one single processor would not have allowed to meet the GREAT10 deadline. Even with a processing time per galaxy of 0.5 seconds, it would have taken about one month to complete process the full GREAT10 dataset. 
The ability to simultaneously run multiple program instances use of parallelism is thus imperative and all programs (denoising, SExtractor wrapper, \emph{gfit}) are written to take advantage of of parallel computers through the Message Passing Interface (MPI) \citep{MPI11995, MPI21998}. When only a few processors are required, the same programs can also run on machines with symmetric multiprocessing (SMP) architecture. It took about 5 days to process the entire GREAT10 challenge images on a 64-processor machine, which corresponds to a processing time between $1$ and $2$ seconds per galaxy. 

The pipeline is implemented in Python, a programming language known for its power, flexibility and short development cycle. The usual standard Python libraries are used, notably: NumPy, SciPy, PyFITS and matplotlib. SciPy is the standard scientific library for Python and most of its functions consist of thin Python wrappers on top of fortran, C and C++ functions. SciPy takes advantage of installed optimized libraries such as LAPACK (Linear Algebra PACKage) library \citep{Anderson1990}. 

\section{Analysis of the gfit GREAT10 results}
\label{section:result analysis}

We summarize and analyze  in this section the main Galaxy Challenge results as far as \emph{gfit} is concerned. An overview of the GREAT10 results for available participating shear measurement methods has already been performed in the GREAT10 Galaxy challenge paper \citep{G10results}. Our objective here is to provide a more detailed analysis of the \emph{gfit} results.

We do not, however, analyze the influence of the pixel-denoising and training calibration schemes applied in \citep{G10results}, which we leave for future investigation.

\subsection{Overall results}
\label{subsection:gfit results}

\begin{figure*}[t]
%\resizebox{\hsize}{!}{\includegraphics[trim=5mm 5mm 5mm 5mm, clip]{fig_gfit_den_all_Q_horiz.pdf}}
\resizebox{\hsize}{!}{\includegraphics[trim=16mm 10mm 22mm 10mm, clip]{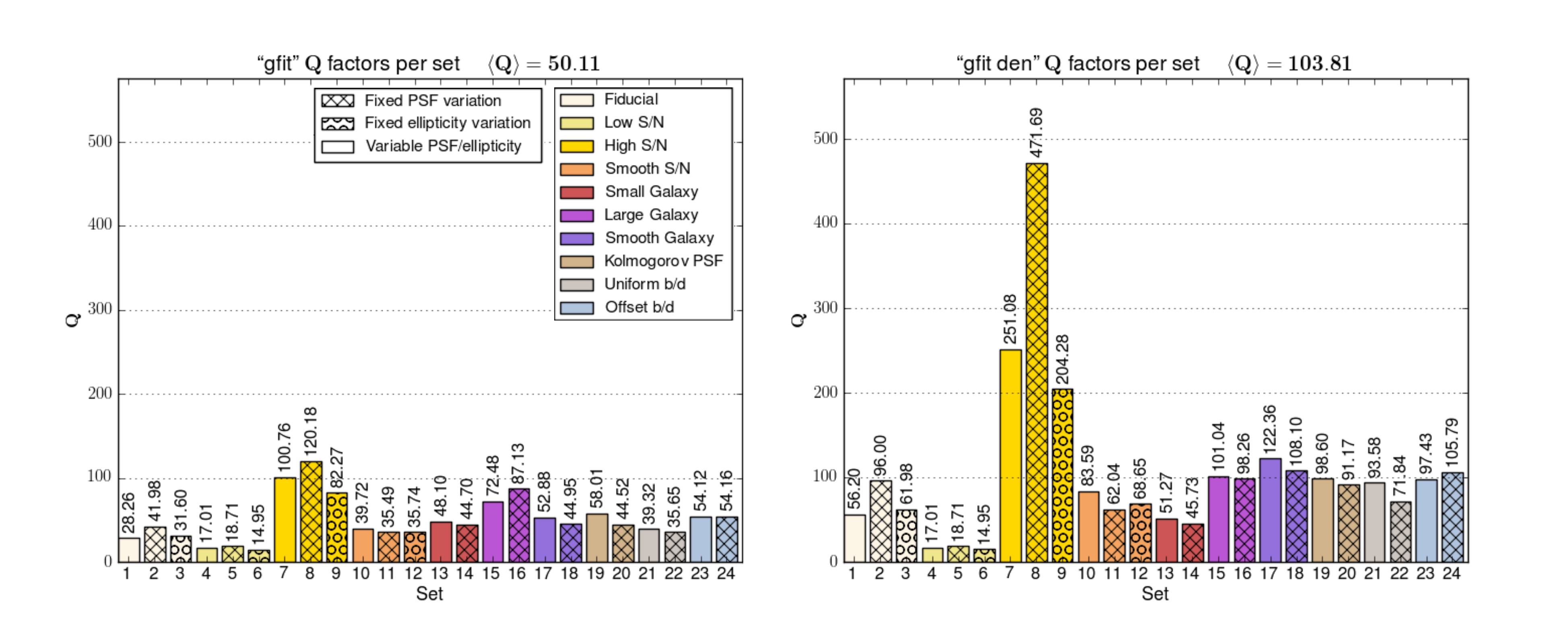}}
\caption{Quality factors per set for ``gfit'' (left) and ``gfit den'' (right). The various colors and patterns in the legend indicate the types of features simulated in the sets. The acronyms $S/N$, $b/d$ respectively refer to the signal to noise ratio and galaxy bulge/disc ratio or offset. The fiducial S/N was $20$ whereas the low $S/N$ and high $S/N$ were respectively set to $S/N=10$ and $S/N=40$. The labels ``Fixed PSF'' and ``Fixed intrinsic ellipticity'' correspond to sets where the PSF and intrinsic galaxy ellipticities were spatially varying across the field but that variation did not change between images within a set. Further details on the structure of the Galaxy challenge dataset and the procedures for calculating the quality factor can be found in in the GREAT10 Galaxy challenge results paper \citet{G10results}.}
\label{fig:Q factors per set vert}
\end{figure*}

%\begin{figure}
%\resizebox{\hsize}{!}{\includegraphics[trim=4mm 5mm 5mm 5mm, clip]{leaderboard_final.pdf}}
%\caption{The list of methods explored in the GREAT10 Galaxy Challenge article \citet{G10results}, in alphabetical order of the method name.}
%\label{fig:leaderboard}
%\end{figure}

The results of the best $12$ methods that participated in the GREAT10 Galaxy Challenge are listed in Table~3 of the \citet{G10results} GREAT10 result paper. That list aggregates results submitted before the official challenge deadline as well as submissions made during the so-called ``Post challenge'', a one-week extension to the competition following the deadline. \\

Two versions of gfit were submitted during the challenge, one named ``gfit den cs'' that included a denoising step using the DWT-Wiener algorithm described in Sect.~\ref{subsection:denoising} and the other, simply named ``gfit'', that did not. The results obtained by both methods are shown in Table~\ref{table:gfit and gfit den cs}. 

The gfit version presented in \citet{G10results} is what we refer to here as "gfit". In addition, we also present "gfit den cs", described in Appendix E5 of \citet{G10results} but whose results were not included in the analysis. The ``gfit'' version is identical except that no denoising was applied to the data before applying the shape measurement algorithm. To simplify, we shorten he name ``gfit den cs'' to ``gfit den'' in the remainder of this article.\\ 

\begin{table}
\caption{Main metrics for ``gfit den'' and ``gfit' .}
\renewcommand{\arraystretch}{1.20}
\label{table:gfit and gfit den cs}
%\begin{center}
\begin{tabular}{llccll}
\hline
Method & $Q$ & $Q_{dn}$ &  $Q_{dn\, \&\, train}$  & $\mathcal{M}/2\times10^{-2}$ & $\sqrt{\mathcal{A}}\times10^{-4}$\\ 
\hline
gfit den &  $103.81$ & 197.88  & $229.19$ & $-2.067$  & $+0.061$ \\
gfit & $50.11$ & $122.74$ & $249.88$ & $+0.583$ &  $+0.057$ \\

\hline
\end{tabular}
\end{table} 

It can be seen from Table~\ref{table:gfit and gfit den cs}  that ``gfit den'' reaches a raw quality factor $Q$ twice as high as that of ``gfit''. This illustrates the gain in accuracy provided by the DWT-Wiener denoising algorithm. This is further analyzed in Sect.~\ref{subsubsection:effect of denoising}. 
When the pixel-level denoising algorithm of \citet{G10results} is applied, the $Q_{dn}$ quality factors of both ``gfit'' and ``gfit den'' are improved by a factor $\sim2$, ``gfit den'' scoring the best  $Q_{dn}$ of all methods ($Q_{dn}=197.88$).
The training calibration further increases both $Q_{dn\, \&\, train}$ quality factors, especially that of ``gfit'' (two-fold increase).

%If we include ``Post challenge'' methods", the ``gfit den'' version obtains the forth best raw $Q$, the best denoised $Q_{dn}$ and the fourth best $Q_{dn \& trained}$, respectively  \mbox{$Q=103.81$}, $Q_{dn}=197.88$, and $Q_{dn \& trained}=229.19$.

We  have also included in Table~\ref{table:gfit and gfit den cs} the average additive and multiplicative biases $\mathcal{A}$ and $\mathcal{M}/2$ over all $24$ sets. Comparing with Table~3 of the \citet{G10results} result paper, we see that ``gfit'' reached the lowest average additive bias \mbox{($\sqrt{\mathcal{A}}=0.057\times10^{-4}$)} and the second lowest average multiplicative bias \mbox{($\mathcal{M}/2=0.583\times10^{-2}$)} of all twelve methods (see also the plot in Figure. 1, page 6 of that paper). 

We stress that, contrary to what is suggested in Sect.~4.4 of \citet{G10results}, the low overall bias of ``gfit'' is intrinsic to the method and \emph{does not} result from the application of a denoising step. The DWT-Wiener algorithm was only used in ``gfit den'', not ``gfit'' and actually, the average multiplicative bias of ``gfit den'' ($\mathcal{M}/2=-2.067\times10^{-2}$) is higher than that of ``gfit''. Moreover, both methods have similar additive biases. Drawing a more refined conclusion about these biases requires an analysis at individual set level, which we perform in Sect.~\ref{subsection:bias analysis per set}.

%The average additive bias of ``gfit den'' is of the same order ($\sqrt{\mathcal{A}}=0.061\times10^{-4}$), but its multiplicative bias increases to $\mathcal{M}/2=-2.067\times10^{-2}$. 

%Table~\ref{table:gfit and gfit den cs metrics S/N} shows how the additive and multiplicative biases of ``gfit''  and ``gfit den'' evolve with the signal to noise ratio. It appears that `gfit den cs'' actually reaches lower biases than ``gfit'' on most sets. A more precise diagnostic requires an analysis at individual set level, which we perform in Sects.~\ref{subsection:accuracy per set} and ~\ref{subsection:bias analysis per set} below.

%Table~\ref{table:gfit and gfit den cs metrics S/N} also shows how the additive and multiplicative biases of ``gfit''  and ``gfit den''  and evolve with the signal to noise ratio. As expected, bias decreases as S/N increases.

%[TODO: 1) compare gfit with the other methods in terms of Q and bias. 2) Process to an analysis per set for gfit, according to different image types.  3) Analyze the gain obtained thanks to denoising (per set)] 

\subsection{Method accuracy}
\label{subsection:accuracy per set}

In this section we use the quality factor as a measure of accuracy and assess the influence of:
\begin{itemize}
\item  Galaxy and PSF characteristics included in the images (size, signal-to-noise, etc.)
\item  Denoising with the DWT-Wiener algorithm
\end{itemize}

%\begin{itemize}
%\item  Signal-to-noise
%\item  Galaxy type
%\item  Galaxy-size
%\item  Bulge/Disk offset
%\item  Turbulence in the PSF
%\end{itemize}

The quality factors scored for each individual image set are plotted on the left-hand side part of Fig.~\ref{fig:Q factors per set vert}, for each \emph{gfit} variant. They are also quoted in Tables~\ref{table:gfit results per set} and \ref{table:gfit den cs results per set}.

Due to an editorial mistake, the shear power spectra attributed to ``gfit'' in the GREAT10 Galaxy Challenge paper, Figure E9, is that from the ``fit2-unfold'' method. The correct picture was made available at the time of publication and can be found \href{http://great.roe.ac.uk/data/galaxy_article_figures/}{here}. We also provide the correct figure in Fig.~\ref{fig:gfit power spectra}. We include the ``gfit den'' power spectrum in Fig.~\ref{fig:gfit den power spectra} as well.\\

%All correct plots for the \citet{G10results} article were made available at the time of publication under \url{http://great.roe.ac.uk/data/galaxy_article_figures/}. We also include the ``gfit den'' power spectrum in Fig.~\ref{fig:gfit den power spectra}.
%%We also include the ``gfit den'' power spectrum in Fig.~\ref{fig:gfit den power spectra}.\\

We focus first on the influence of galaxy and PSF features on accuracy. We leave aside the effects of DWT-Wiener denoising for now and thus base our analysis on the results of the ``gfit'' variant which is devoid of built-in denoising scheme. 
%We also only consider the ``raw'' quality factor ($Q$) since the other quality factors account for pixel-level denoising and training. 
The data of interest are summarized in Figs.~\ref{fig:Q factors per set vert} and \ref{fig:gfit power spectra} and in Tables~\ref{table:gfit results per set}, \ref{table:gfit den cs results per set}.

\subsubsection{Influence of Galaxy and PSF characteristics}
\label{subsubsection:effect of features}

\begin{itemize}
\item \textbf{Influence of signal-to-noise ratio}: this is best reflected on the results of the ``gfit'' variant, since it has no built-in noise correction scheme. All that was done regarding noise was to cut out galaxy and PSF postage stamps to lower dimensions: $12\times12$ pixels for the PSF and $24\times24$ pixels for galaxies.  As expected, higher S/N yields higher Q factors: we observe a roughly linear progression with a $\sim$ two-fold increase from low $S/N=10$ to fiducial $S/N=20$ and a $\sim$ three-fold increase from low $S/N=10$ to higher $S/N=40$ (see sets $4$ to $12$ in Fig.~\ref{fig:Q factors per set vert}).\\

%no noise correction algorithm was applied to the ``gfit'' variant. All that was done was reduce the postage stamp size of the galaxies and PSF in an at   as expected, higher S/N yields higher Q factors: a roughly linear progression with a $\sim$ two-fold increase from low $S/N=10$ to fiducial $S/N=20$ and a $\sim$ three-fold increase from low $S/N=10$ to higher $S/N=40$.
%\item {Influence of galaxy size}: we observe that accuracy decrease smaller galaxies 
%\item {Influence of signal-to-noise ratio}: no noise correction algorithm was applied to the ``gfit'' variant. All that was done was reduce the postage stamp size of the galaxies and PSF in an at   as expected, higher S/N yields higher Q factors: a roughly linear progression with a $\sim$ two-fold increase from low $S/N=10$ to fiducial $S/N=20$ and a $\sim$ three-fold increase from low $S/N=10$ to higher $S/N=40$.

\item \textbf{Influence of galaxy size}: as seen from the results of sets $13$ to $18$ in in Fig.~\ref{fig:Q factors per set vert}), ``gfit'' seems quite sensitive to galaxy size, the worst Q factors being obtained on smaller galaxies and the best on the larger ones, with a factor $\sim$~2 difference. The use of postage stamp cut-outs of identical dimensions, regardless of the actual FWHM of the galaxy they contain could be responsible for this effect, as cut-outs with smaller galaxies are likely to be more noise-dominated than those with larger objects. It may also be that the minimizer is less accurate on smaller objects. ``Smooth'' galaxies have sizes varying according to a Rayleigh distribution \citep{G10results}, so it is not surprising that the Q factors of the corresponding sets 10, 11, 12 take values in between those of small and large galaxies.\\

\item  \textbf{Influence of bulge/disc distribution and offset}: having varying b/d ratios (sets 21, 22) seems to decrease accuracy down to the level of small-size galaxies (sets 13, 14). On the other hand, introducing a non-zero 50/50 bulge-to-disc (b/d) offset (sets 21, 22) tends to yield slightly higher accuracy compared to ``smooth'' galaxies (sets 17, 18). So it seems the single-component galaxy model (see Sect.~\ref{subsection:galaxy model}) of ``gfit'' is more sensitive to b/d ratio than to b/d separation. The use of a more sophisticated galaxy model would certainly yield a small gain for some types of galaxies but at the price of an additional computational time for model fitting. On real data, the importance of the b/d distribution and offset will probably depend on the available galaxy sample. \\ 

\item \textbf{Influence of turbulence}: The raw Q factor plot in Fig.~\ref{fig:Q factors per set vert} does not show a strong impact from the inclusion of a Kolmogorov power spectrum in the PSF ellipticities. We note however a higher score on the fiducial turbulent set compared to the non-turbulent one for no obvious reason. This seems counter intuitive as PSF turbulence usually degrades accuracy and this phenomenon may actually not be related to turbulence.\\

\item \textbf{Influence of spatial variability between images}: as described in \citet{G10results}, the Galaxy challenge data are divided into so-called ``Single epoch'', ``Multi-epoch'' and ``Stable single epoch'' depending on whether the intrinsic ellipticities and PSF keep the same or change their spatial distribution between images in a set (see Table~\ref{table:spatial variability}). It seems that having fixed instead of variable intrinsic ellipticities (``Multi-epoch'' sets) slightly decreases accuracy. Apart from this, no clear trend really stands out from the results and it is not clear whether the difference in accuracy is due to the type of spatial variability used or to the specific sample of images chosen for a set. This topic nevertheless deserves to be investigated further in a separate work.

\end{itemize}

% --- Mutiplicative Bias Plot %
\begin{figure*}[t]
%\resizebox{\hsize}{!}{\includegraphics[trim=5mm 5mm 5mm 5mm, clip]{fig_gfit_den_all_Q_horiz.pdf}}
\resizebox{\hsize}{!}{\includegraphics[trim=11mm 10mm 22mm 10mm, clip]{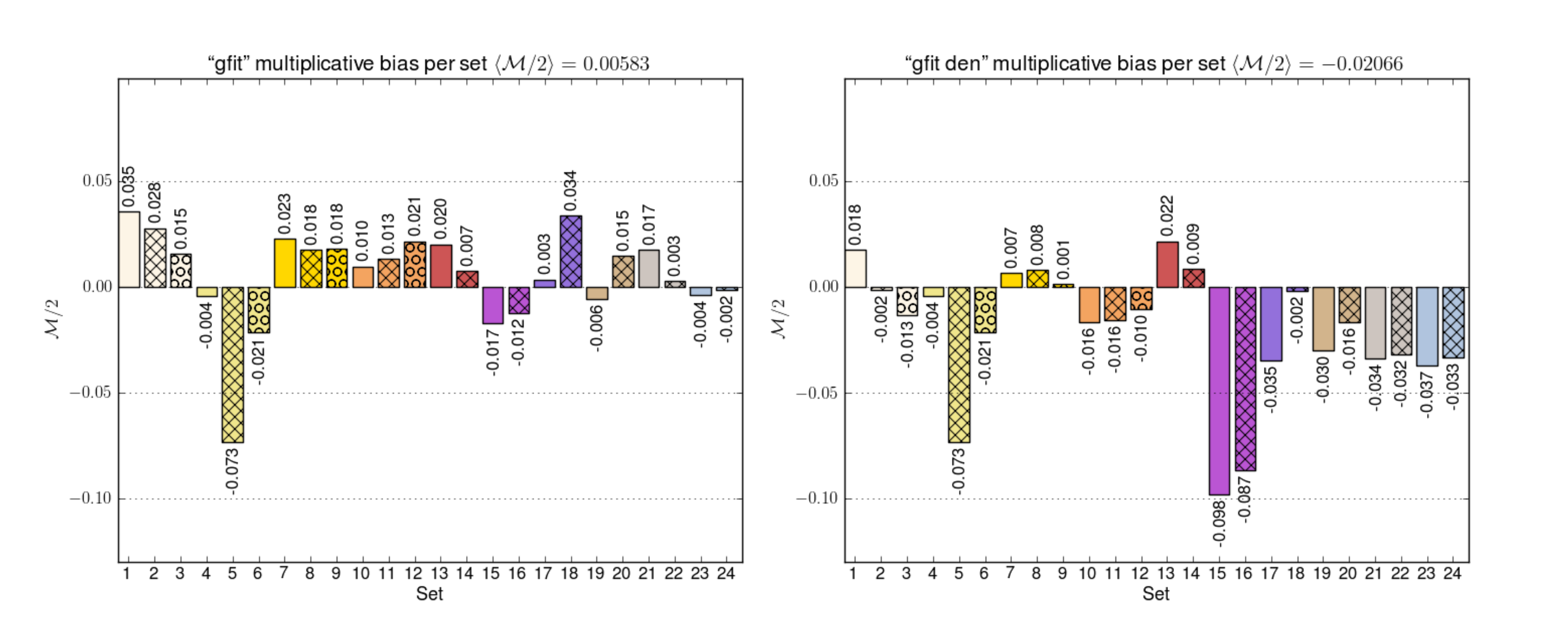}}
%\resizebox{\hsize}{!}{\includegraphics[trim=16mm 10mm 22mm 10mm, clip]{fig_multiplicative_bias.pdf}}
\caption{Multiplicative bias $\mathcal{M}/2$ per set. The legend patterns and colors are identical to those of Fig.~\ref{fig:Q factors per set vert}}
\label{fig:multiplicative bias per set}
\end{figure*}

% --- Additive Bias Plot %
\begin{figure*}[t]
%\resizebox{\hsize}{!}{\includegraphics[trim=5mm 5mm 5mm 5mm, clip]{fig_gfit_den_all_Q_horiz.pdf}}
\resizebox{\hsize}{!}{\includegraphics[trim=11mm 10mm 22mm 10mm, clip]{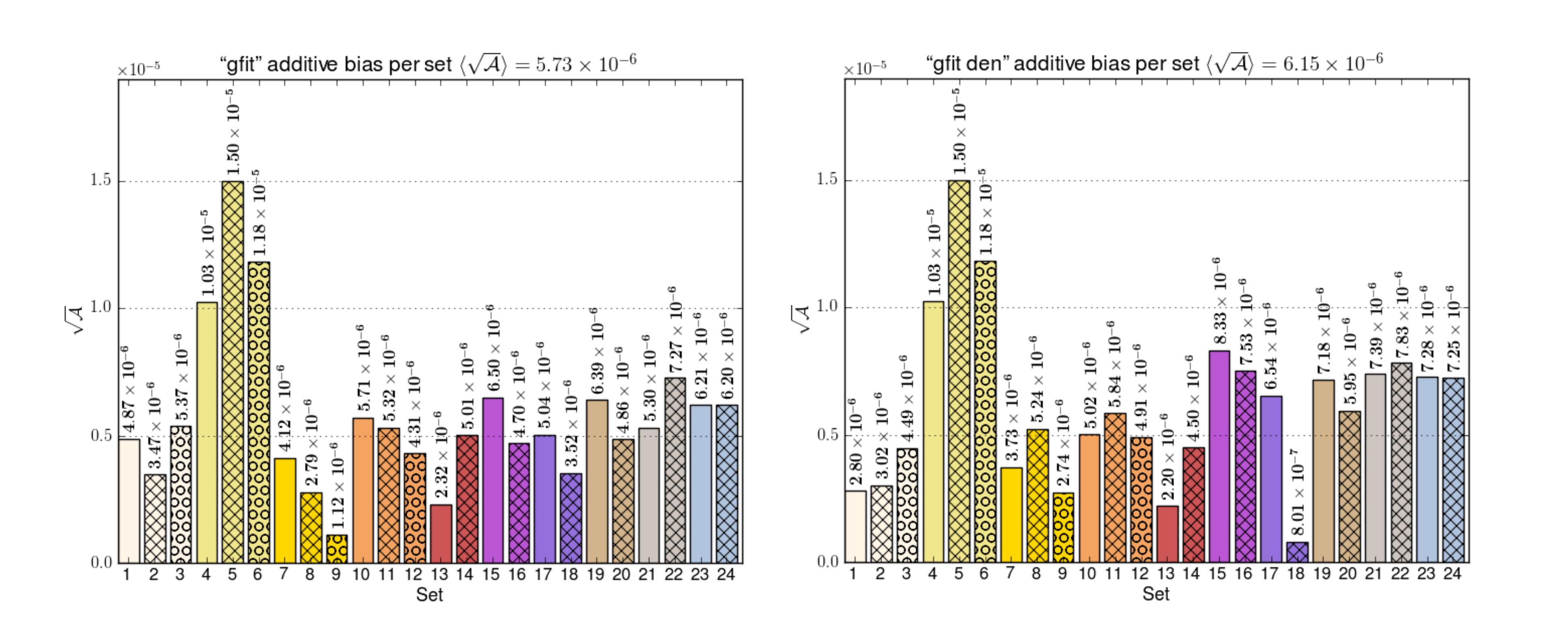}}
\caption{Additive bias $\sqrt{\mathcal{A}}$ per set. The legend patterns and colors are identical to those of Fig.~\ref{fig:Q factors per set vert}}
\label{fig:additive bias per set}
\end{figure*}

%\begin{itemize}
%\item Influence of galaxy and PSF features on accuracy
%\item Influence of roughly on accuracy
%\end{itemize}

\subsubsection{Effect of denoising on accuracy}
\label{subsubsection:effect of denoising}

We discuss in this section the effect on accuracy of the application of the DWT-Wiener denoising scheme described in Sect.~\ref{subsection:denoising}.
The influence of denoising can be clearly observed by comparing the plot of ``gfit den'' (left) with that of ``gfit'' in Fig.~\ref{fig:Q factors per set vert} (right). The corresponding scores are also listed in Tables~\ref{table:gfit results per set} and \ref{table:gfit den cs results per set}.

We find an average two-fold increase in accuracy, the effect being stronger on high S/N images and larger galaxies. Sets with small galaxies are only slightly improved, however. We also notice that the plots of ``gfit'' and ``gfit den'' show identical quality factors for low S/N sets 4 to 6. Further investigation showed that denoising was, by mistake, not applied on those sets. This would probably have improved the overall Q factor of ``gfit den''.

These results strongly suggest that the DWT-Wiener algorithm really improves the overall accuracy on galaxy shape measurement. We also note that denoising does not alter the quality factor hierarchy between sets: the sets with best scores in the ``gfit'' plots remain the same in the ``gfit den'' plot.

\subsection{Bias analysis}
\label{subsection:bias analysis per set}

We investigate in this section how multiplicative and additive biases are affected by galaxy properties and the use of denoising.

\subsubsection{Influence of Galaxy and PSF characteristics}
\label{subsubsection:effect of features on bias}

To complement the results of Table \ref{table:gfit and gfit den cs} relative to bias, we have plotted in Figs.~\ref{fig:multiplicative bias per set} and \ref{fig:additive bias per set} the multiplicative and additive biases of each set.
As noted in Sec.~\ref{subsection:gfit results}, ``gfit den'' reached the lowest average additive bias and the second lowest average multiplicative bias of all 12 twelve competing methods.

Focusing on the multiplicative bias and leaving aside the effect of denoising for now, we can make a few observations from the left-hand side plot of Figs.~\ref{fig:multiplicative bias per set}.

\begin{itemize}
%\item \textbf{Highest and lowest multiplicative bias}: all multiplicative bias values, except one, lie below $\mathcal{M}/2=3.5\times10^{-2}$ in absolute value with a tendency to be negative. Leaving aside the effects of spatial variability, the larger biases are found on fiducial set 1 and the set with small galaxies (set 13). The smallest biases are obtained on images with large galaxies and $50/50$ bulge-to-disc ratio galaxies.\\

%\item \textbf{Highest and lowest multiplicative bias}: if we only take into account sets with variable PSF and ellipticities (1, 4, 7, 10, 13, 17, 19, 23, 25), multiplicative bias values, except one, lie below $\mathcal{M}/2=3.5\times10^{-2}$ in absolute value. For the same sets, the smallest biases are obtained on images with large galaxies (17) and galaxies with $50/50$ bulge-to-disc (b/d) ratios with offset (set 25). On the other hand, if we consider set with fixed PSF and ellipticity variations (2, 3, 5, 6, 8, 9, 11, 12, 14, 16, 18, 20, 24, 26) the largest biases are reached on low S/N images (sets 5, 6) and large galaxies and the smallest on images with uniform and offset b/d.\\

%``Single epoch'', ``Multi-epoch'' and ``Stable single epoch''

\item \textbf{Highest and lowest multiplicative bias}: all multiplicative bias values, except one (set 5), lie below $\mathcal{M}/2=3.5\times10^{-2}$ in absolute value. The largest biases are found on fiducial set 1 (fixed PSF \& variable ellipticities), set 5 (low S/N, fixed PSF) and set 18 (smooth galaxies, fixed PSF). The smallest biases are obtained on set 4 (low S/N, fixed PSF \& variable ellipticities), set 17 (smooth galaxies, fixed PSF) and sets 22 to 24 (galaxies with varying b/d ratios and fixed PSF, galaxies with non-zero b/d offset).\\

\item \textbf{Influence of signal-to-noise ratio}: the results from sets 1 and 7 suggest, as expected, that multiplicative bias decreases with S/N. The low S/N set 4, however, shows a very small bias. It may be that the true bias is large and positive but that it was offset by e.g. a negative bias due to the contribution of large galaxies in that particular set. The relatively large bias of set 5 compared to sets 4 and 6 may also be an consequence of the PSF having a fixed variation pattern in that set.\\

\item \textbf{Influence of galaxy size}: small galaxies (sets 13, 14) and large galaxies (sets 15, 16) have comparable biases in absolute value, smaller galaxies having a positive bias and larger ones a negative bias. A lower bias is reached on ``Smooth'' galaxy images (set 17), likely because the negative and positive biases of small and large galaxies respectively compensate each others. The bias in set 18 may also have been artificially amplified by the fixed variation pattern of the PSF in that set.\\

\item \textbf{Influence of bulge/disc distribution and offset}: letting the b/d ratios of galaxy vary within a set tends to yield a multiplicative bias comparable to that of small-size galaxies, probably causing the decrease in accuracy mentioned in Sect.~\ref{subsection:accuracy per set}. In contrast, the biases associated with a non-zero $50/50$ b/d offset are among the lowest. As noted earlier, the underlying single-component S\'ersic-based galaxy model of ``gfit'' (see Sect.~\ref{subsection:galaxy model}) seems handle quite well profiles with an off-centered bulge and disk.\\ 

\item \textbf{Influence of turbulence}: the introduction of PSF with turbulent ellipticities (set 19) does not induce a particularly significant bias compared to the average.\\

\item \textbf{Influence of spatial variability between images}: we note that having ``Fixed PSF'' and ``Fixed intrinsic ellipticity'' significantly alters the multiplicative bias, especially on low S/N and smooth galaxy sets. Because all images have the same spatial variation within a set, its is likely that the bias of one image is just amplified as many times as there are images (i.e. 200 times). The effect is particularly strong on image with low S/N (set 5) and smooth galaxies (set 18).
\end{itemize}
As far as additive bias is concerned, we note the following trend:

\begin{itemize}
\item \textbf{Highest and lowest additive bias}: ``gfit'' obtains an additive bias $\sqrt{\mathcal{A}} \lesssim~10^{-5}$ on all types of sets, except on the low S/N ones. All values are positive. The lowest bias is reached on small galaxies (sets 13, 14) and the largest on low S/N images (sets $4$ to $6$). \\

\item \textbf{Influence of signal-to-noise ratio}: as for multiplicative bias, we find a trend toward higher biases for lower S/N (sets 1 to 9). \\

\item \textbf{Influence of turbulence}: the introduction of a Kolmogorov power spectrum in PSF ellipticity induces a slightly greater additive bias than average (see set 19). \\

\item \textbf{Influence of galaxy size}: as for multiplicative bias, we note a tendency of smaller galaxies to have a higher additive bias, as reflected by the values for sets $15$ to $18$. \\

\item \textbf{Influence of bulge/disc distribution and offset}: the additive bias appears larger on galaxies with varying b/d ratios (sets 21, 22) and non-zero b/d offset (sets 23, 24).

\end{itemize}

\subsubsection{Effect of denoising on bias}
\label{subsubsection:effect of denoising on bias}

Comparing the plots for ``gfit'' and ``gfit den'' in Figs.~\ref{fig:multiplicative bias per set} and \ref{fig:additive bias per set}, we find that the additive bias does not change significantly, keeping about the same bias values per set.
In contrast, the structure of the multiplicative plots is significantly altered. %We summarize our main observations below. 

As seen on the ``gfit den'' plot, denoising tend to introduce some amount of negative multiplicative bias on all sets. Although the amount of bias on fiducial sets remains roughly the same in absolute value, the DWT-Wiener algorithm clearly impacts the multiplicative bias relative to galaxy size, b/d ratio and turbulence. 

As regards galaxy size, even though the multiplicative bias on small galaxies is almost unchanged, that on large galaxies increases about six-fold. Because of that, the bias on ``Smooth'' galaxies, which also includes large galaxies, also increases. 

DWT-Wiener denoising seems to also improves the resolution of the bulge and disk components, so that ``gfit'' has more difficulty fitting its underlying single-component S\'ersic model to images with ``uniform'' and ``offset'' b/d. The effect is stronger on larger galaxies, causing a $\gtrsim$ tenfold multiplicative bias increase. 

Lastly, the introduction of PSF turbulence result in a $\sim$ five-fold multiplicative bias degradation.

Despite the degradation of multiplicative bias on some sets, the accuracy of shape measurements increases two-fold as shown by the corresponding gain in Q factor. We also note that denoising improves the results on sets that already have a high S/N. All in all, the use of denoising is thus clearly beneficial.

\section{Conclusions}
\label{section:conclusions}

%We have described in this paper \emph{gfit}, a simple model-fitting shape measurement method (Sect.~\ref{section:gfit method description}), and provided an analysis of the results recently obtained in the GREAT10 Galaxy Challenge (Sect.~\ref{section:applying gift on g10 data}). \textbf{TODO: include a summary of gfit here}
%We have described in this paper the \emph{gfit} shape measurement method (Sect.~\ref{section:gfit method description}) and provided an analysis of the results recently obtained in the GREAT10 Galaxy Challenge (Sect.~\ref{section:applying gift on g10 data}).\\

We have described in this paper the \emph{gfit} shape measurement method, a model-fitting based on a simple S\'ersic galaxy model (Sect.~\ref{section:gfit method description}). The method uses a custom-developed minimizer based on a ``coordinate descent'' algorithm that finds a local minimum with the lowest $\chi ^2$ of the residuals between true and modeled galaxy.\\

%As described in Sect.~\ref{section:gfit method description}, \emph{gfit} is a model-fitting based on a simple S\'ersic galaxy model. The method uses a custom-developed minimizer based on an "adaptive cyclic coordinate descent algorithm" that find a local minimum with the lowest $\chi ^2$ of the residuals between true and modeled galaxy.\\

We have also performed an analysis of our results in the GREAT10 Galaxy Challenge (Sect.~\ref{section:applying gift on g10 data}). We participated in the competition with two \emph{gfit} variants: ``gfit den'', which applied a denoising step before performing model-fitting, and ``gfit'', which did not use denoising. The noise removal technique employed is DWT-Wiener, a wavelet-based, shape-preserving algorithm particularly suitable for shape measurement. (see Sect.~\ref{subsection:denoising}).\\

%We participated on the Galaxy Challenge with two \emph{gfit} variants: ``gfit den'', which applied a denoising step before performing model-fitting, and``gfit'', which did not use denoising. The denoising technique employed is DWT-Wiener, a wavelet-based, shape-preserving algorithm developed at EPFL and particularly suitable for shape measurement. (see Sect.~\ref{subsection:denoising}).

%It is interesting to see that, despite the simplicity of the galaxy model used, \emph{gfit} established itself as one of the four top-performing methods in the Galaxy Challenge, both in terms of accuracy and bias. This raises the question of how important having a realistic galaxy model really matters when measuring galaxy shapes from real data. Providing more clues on that question is one of the objectives of the forthcoming GREAT3 challenge.\\

%It is interesting to see that, despite the simplicity of the galaxy model used, both variants of \emph{gfit} obtained very competitive results, both in terms of accuracy and bias. This raises the question on how important the accuracy of the galaxy model really matters when measuring galaxy shapes from real data.\\

We highlight below the main conclusions of your analysis.

\begin{itemize}
\item \textbf{Method accuracy}: accuracy improves significantly as S/N gets higher and galaxy size larger. The underlying simple S\'ersic-based galaxy model of gfit has more difficulty handling galaxies with a non-zero $50/50$ offset between bulge and disc. The inclusion of Kolomogorov turbulence in ellipticities is not seen to yield a significant change in accuracy.\\

\item  \textbf{Additive and multiplicative bias}: the non-denoised ``gfit'' variant reached the lowest average additive bias and second lowest average multiplicative bias of all twelve competing methods. Both additive and multiplicative bias tend to be larger on galaxies with high S/N, smaller size and galaxies with b/d ratio differing from $50/50$. \\

\item \textbf{Impact of denoising}: the application of the DWT-Wiener noise removal algorithm yields a two-fold improvement in accuracy (Q factor) despite significantly degrading the multiplicative bias on galaxies with a high S/N, small size and significant bulge/disk ratio and separation. 

\end{itemize}

It is interesting to see that, despite the simplicity of the galaxy model used, its results in the Galaxy Challenge established \emph{gfit} as one of the four top-performing methods, both in terms of accuracy and bias. Given that the results were obtained on simulated data, this raises the question of how important having a realistic galaxy model really matters when measuring galaxy shapes from real data. Providing more clues on this question is one of the objectives of the forthcoming GREAT3 challenge.\\

%It is interesting to see that, despite the simplicity of the galaxy model used, its results in the Galaxy Challenge established \emph{gfit} as one of the four top-performing methods, both in terms of accuracy and bias. This raises the question of how important having a realistic galaxy model really matters when measuring galaxy shapes from real data. Providing more clues on that question is one of the objectives of the forthcoming GREAT3 challenge.\\

\begin{acknowledgements}
This work is supported by the Swiss National Science Foundation (SNSF). Many thanks to Tom Kitching for his help and for sharing the shear analysis code. We also thank the GREAT10 Coordination Team for organizing this stimulating challenge. GREAT10 was sponsored by a EU FP7 PASCAL 2 challenge grant. We also acknowledge support from the International Space Science Institute (ISSI) in Bern, where some of this research has been discussed.
\end{acknowledgements}

\bibliographystyle{aa}
\bibliography{../Articles}

\clearpage

\begin{appendix}
\label{appendix}

\section{Accuracy and bias per set}
\label{section:appendix results per set}

Tables~\ref{table:gfit results per set} and \ref{table:gfit den cs results per set} respectively quote the actual quality factor and bias values reached by the non-denoised and denoised variants of the \emph{gfit} shape measurement method.

\definecolor{S/N10}{RGB}{255,236,139}
\definecolor{S/N20}{RGB}{193,255,193}
\definecolor{S/N40}{RGB}{187,255,255}
\definecolor{ALL}{RGB}{224,238,238}

\begin{table}
\caption{``gfit'': results per set. Sets with S/N 10, 40 are respectively highlighted in orange and blue. Fiducial sets with S/N 20 are represented in green and all remaining sets also have a S/N of 20.}
\renewcommand{\arraystretch}{1.30}
\label{table:gfit results per set}
\begin{tabular}{rrrrlll}
\hline
Set & $Q$ & $Q_{dn}$ &  $Q_{dn\, \&\, train}$  & $\mathcal{M}/2\times10^{-2}$ & $\sqrt{\mathcal{A}}\times10^{-4}$\\ 
\hline
\cellcolor{S/N20}$1$&\cellcolor{S/N20}$28.26$&\cellcolor{S/N20}$50.44$&\cellcolor{S/N20}$205.57$&\cellcolor{S/N20}$+3.542$&\cellcolor{S/N20}$+0.0487$\\
\cellcolor{S/N20}$2$&\cellcolor{S/N20}$41.98$&\cellcolor{S/N20}$78.50$&\cellcolor{S/N20}$352.54$&\cellcolor{S/N20}$+2.770$&\cellcolor{S/N20}$+0.0347$\\
\cellcolor{S/N20}$3$&\cellcolor{S/N20}$31.60$&\cellcolor{S/N20}$77.86$&\cellcolor{S/N20}$411.57$&\cellcolor{S/N20}$+1.539$&\cellcolor{S/N20}$+0.054$\\
\cellcolor{S/N10}$4$&\cellcolor{S/N10}$17.01$&\cellcolor{S/N10}$38.11$&\cellcolor{S/N10}$114.55$&\cellcolor{S/N10}$-0.432$&\cellcolor{S/N10}$+0.103$\\
\cellcolor{S/N10}$5$&\cellcolor{S/N10}$18.71$&\cellcolor{S/N10}$34.41$&\cellcolor{S/N10}$89.59$&\cellcolor{S/N10}$-7.345$&\cellcolor{S/N10}$+0.150$\\
\cellcolor{S/N10}$6$&\cellcolor{S/N10}$14.95$&\cellcolor{S/N10}$31.32$&\cellcolor{S/N10}$63.61$&\cellcolor{S/N10}$-2.136$&\cellcolor{S/N10}$+0.118$\\
\cellcolor{S/N40}$7$&\cellcolor{S/N40}$100.75$&\cellcolor{S/N40}$450.35$&\cellcolor{S/N40}$111.06$&\cellcolor{S/N40}$+2.270$&\cellcolor{S/N40}$+0.041$\\
\cellcolor{S/N40}$8$&\cellcolor{S/N40}$120.18$&\cellcolor{S/N40}$308.43$&\cellcolor{S/N40}$143.61$&\cellcolor{S/N40}$+0.176$&\cellcolor{S/N40}$+0.028$\\
\cellcolor{S/N40}$9$&\cellcolor{S/N40}$82.27$&\cellcolor{S/N40}$185.17$&\cellcolor{S/N40}$184.27$&\cellcolor{S/N40}$+1.780$&\cellcolor{S/N40}$+0.011$\\
$10$&$39.72$&$89.40$&$319.88$&$+0.964$&$+0.057$\\
$11$&$35.49$&$81.86$&$357.68$&$+1.327$&$+0.053$\\
$12$&$35.74$&$86.16$&$348.86$&$+2.141$&$+0.043$\\
$13$&$48.10$&$121.62$&$215.05$&$+2.010$&$+0.023$\\
$14$&$44.70$&$114.73$&$254.17$&$+0.740$&$+0.050$\\
$15$&$72.48$&$155.84$&$193.04$&$-1.732$&$+0.065$\\
$16$&$87.13$&$245.81$&$135.62$&$-1.238$&$+0.047$\\
$17$&$52.88$&$125.30$&$246.64$&$+0.334$&$+0.050$\\
$18$&$44.95$&$117.64$&$180.68$&$+3.366$&$+0.035$\\
$19$&$58.01$&$113.55$&$284.16$&$-0.593$&$+0.064$\\
$20$&$44.52$&$90.47$&$436.67$&$+1.465$&$+0.049$\\
$21$&$39.32$&$73.23$&$291.54$&$+1.732$&$+0.053$\\
$22$&$35.65$&$67.33$&$345.03$&$+0.271$&$+0.073$\\
$23$&$54.12$&$105.81$&$339.60$&$-0.382$&$+0.062$\\
$24$&$54.15$&$102.45$&$372.16$&$-0.164$&$+0.062$\\
\hline
\textbf{All}&$\mathbf{50.11}$&$\mathbf{122.74}$&$\mathbf{249.88}$&$\mathbf{+0.583}$&$\mathbf{+0.057}$\\
%\cellcolor{ALL}All&\cellcolor{ALL}$50.11$&\cellcolor{ALL}$122.74$&\cellcolor{ALL}$249.88$&\cellcolor{ALL}$+0.583$&\cellcolor{ALL}$+0.057$\\
\hline
\end{tabular}
\end{table}

\begin{table}[!b]
\caption{``gfit den'': results per set. Sets with S/N 10, 40 are respectively highlighted in orange and blue. Fiducial sets with S/N 20 are represented in green and all remaining sets also have a S/N of 20.}
\renewcommand{\arraystretch}{1.30}
\label{table:gfit den cs results per set}
\begin{tabular}{rrrrlll}
\hline
Set & $Q$ & $Q_{dn}$ &  $Q_{dn\, \&\, train}$  & $\mathcal{M}/2\times10^{-2}$ & $\sqrt{\mathcal{A}}\times10^{-4}$\\ 
\hline
\cellcolor{S/N20}1&\cellcolor{S/N20}$56.20$ &\cellcolor{S/N20}$182.89$ &\cellcolor{S/N20}$269.98$ &\cellcolor{S/N20}$+1.761$ &\cellcolor{S/N20}$+0.028$ \\
\cellcolor{S/N20}2 &\cellcolor{S/N20}$96.00$ &\cellcolor{S/N20}$316.69$ &\cellcolor{S/N20}$291.60$ &\cellcolor{S/N20}$-0.154$ &\cellcolor{S/N20}$+0.030$ \\
\cellcolor{S/N20}3 &\cellcolor{S/N20}$61.98$ &\cellcolor{S/N20}$271.56$ &\cellcolor{S/N20}$272.04$ &\cellcolor{S/N20}$-1.321$ &\cellcolor{S/N20}$+0.045$ \\
\cellcolor{S/N10}4 &\cellcolor{S/N10}$17.01$ &\cellcolor{S/N10}$38.11$\cellcolor{S/N10}& $59.06$\cellcolor{S/N10}& $-0.432$\cellcolor{S/N10}&\cellcolor{S/N10}$+0.103$ \\
\cellcolor{S/N10}5 &\cellcolor{S/N10}$18.71$ &\cellcolor{S/N10}$34.41$ &\cellcolor{S/N10}$54.09$ &\cellcolor{S/N10}$-7.345$ &\cellcolor{S/N10}$+0.150$ \\
\cellcolor{S/N10}6 &\cellcolor{S/N10}$14.95$ &\cellcolor{S/N10}$31.32$ &\cellcolor{S/N10}$45.34$ &\cellcolor{S/N10}$-2.136$ &\cellcolor{S/N10}$+0.118$ \\
\cellcolor{S/N40}7 &\cellcolor{S/N40}$251.08$ &\cellcolor{S/N40}$518.17$ &\cellcolor{S/N40}$143.08$ &\cellcolor{S/N40}$+0.671$ &\cellcolor{S/N40}$+0.037$ \\
\cellcolor{S/N40}8 &\cellcolor{S/N40}$471.69$ &\cellcolor{S/N40}$203.29$ &\cellcolor{S/N40}$108.09$ &\cellcolor{S/N40}$+0.821$ &\cellcolor{S/N40}$+0.052$ \\
\cellcolor{S/N40}9 &\cellcolor{S/N40}$204.28$ &\cellcolor{S/N40}$503.64$ &\cellcolor{S/N40}$156.95$ &\cellcolor{S/N40}$+0.137$ &\cellcolor{S/N40}$+0.027$ \\
10 &  $83.59$ & $303.87$ & $269.16$ & $-1.647$ & $+0.050$ \\
11 &  $62.04$ & $162.31$ & $285.93$ & $-1.556$ & $+0.058$ \\
12 &  $68.65$ & $258.49$ & $412.10$ & $-1.022$ & $+0.049$ \\
13 &  $51.27$ & $165.24$ & $259.73$ & $+2.166$ & $+0.022$ \\
14 &  $45.73$ & $127.28$ & $258.15$ & $+0.863$ & $+0.045$ \\
15 &  $101.04$ & $100.83$ & $85.12$ & $-9.790$ & $+0.083$ \\
16 &  $98.26$ & $101.14$ & $94.13$ & $-8.678$ & $+0.075$ \\
17 &  $122.36$ & $207.82$ & $266.86$ & $-3.468$ & $+0.065$ \\ 
18 &  $108.10$ & $264.72$ & $204.89$ & $-0.202$ & $+0.008$ \\
19 &  $98.59$ & $154.86$ & $286.79$ & $-0.301$ & $+0.072$ \\
20 &  $91.17$ & $198.34$ & $548.31$ & $-1.640$ & $+0.059$ \\
21 &  $93.58$ & $158.18$ & $257.46$ & $-3.363$ & $+0.074$ \\
22 &  $71.84$ & $137.44$ & $327.12$ & $-3.189$ & $+0.078$ \\
23 &  $97.43$ & $144.90$ & $265.60$ & $-0.037$ & $+0.073$ \\
24 &  $105.79$ & $163.57$ & $278.86$ & $-3.339$ & $+0.072$ \\
\hline
%\cellcolor{ALL}All&\cellcolor{ALL}$103.81$ &\cellcolor{ALL}$197.88$ &\cellcolor{ALL}$229.19$ &\cellcolor{ALL}$-2.067$ &\cellcolor{ALL}$+0.061$ \\
\textbf{All}&$\mathbf{103.81}$ &$\mathbf{197.88}$ &$\mathbf{229.19}$ &$\mathbf{-2.067}$ &$\mathbf{+0.061}$ \\
\hline
\end{tabular}
\end{table}

\section{Sear power spectra}
\label{subsection:appendix power spectra}

\begin{figure*}[b]
\begin{center}
\resizebox{\hsize}{!}{\includegraphics[trim=15mm 110mm 22mm 80mm, clip]{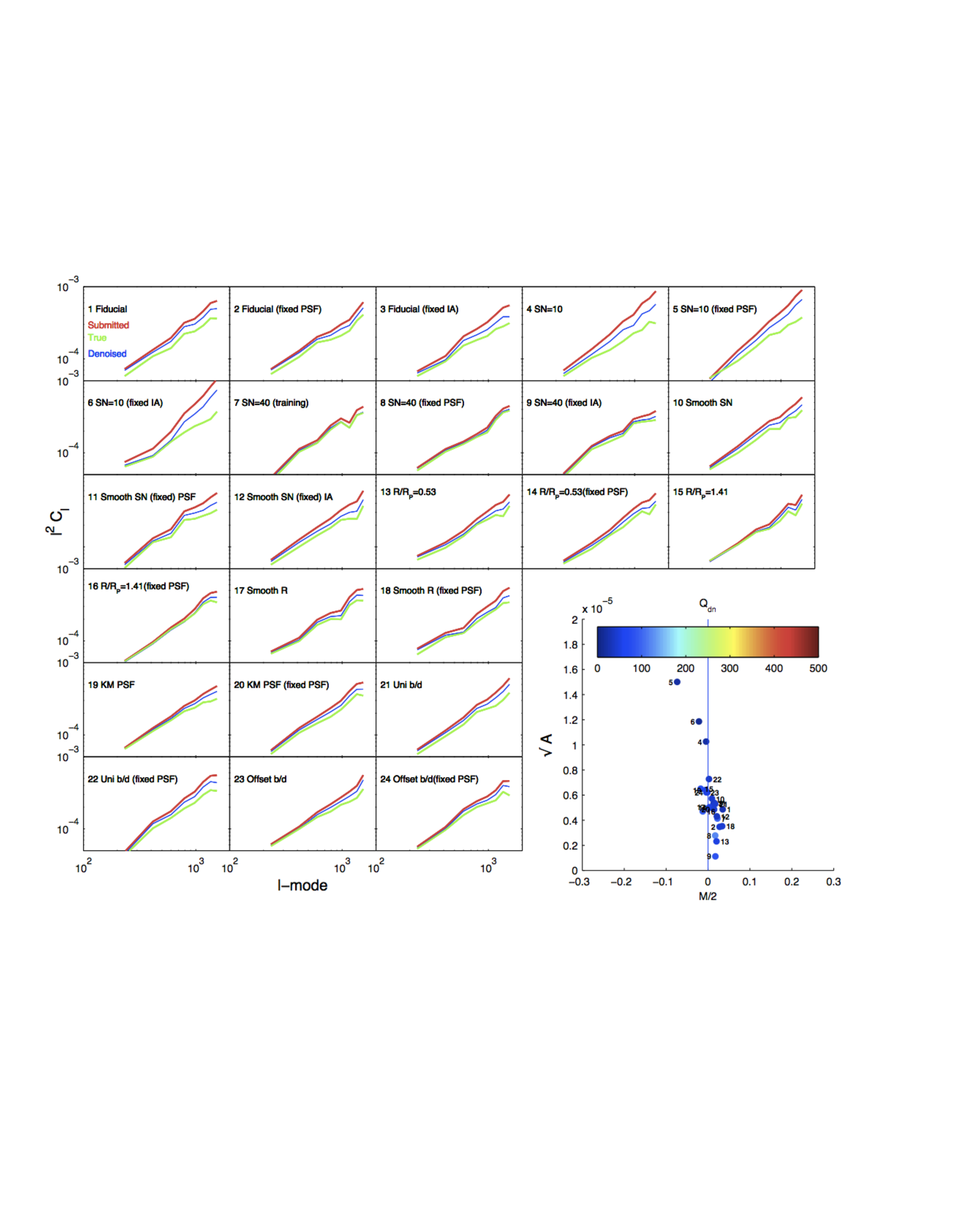}}
\caption{The shear power spectra of the ``gfit'' variant over the 24 sets of the Galaxy challenge. The red lines denote the estimated shear power spectra while the green lines represents the true shear power. The blue lines indicate the spectra after application of pixel-level denoising (not DWT-Wiener denoising). The code run to plot these power spectra is identical to that used in the \citep{G10results} paper. Note that the power spectra attributed to ``gfit'' in the Figure E9 of \citep{G10results} are actually those of the ``fit2-unfold'' method.}
\label{fig:gfit power spectra}
\end{center}
\end{figure*}

\begin{figure*}[b]
\begin{center}
\resizebox{\hsize}{!}{\includegraphics[trim=15mm 110mm 22mm 80mm, clip]{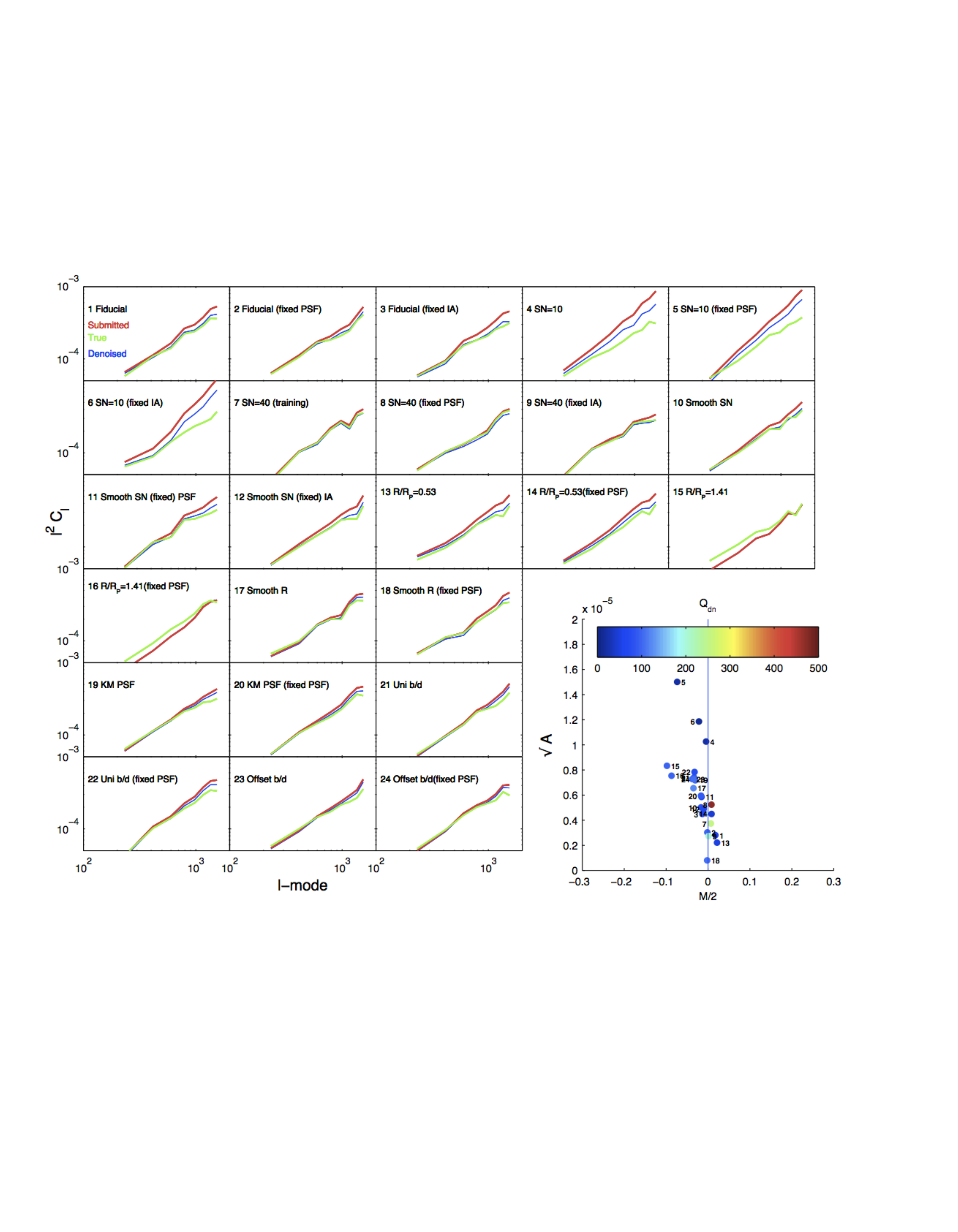}}
\caption{The shear power spectra  of the ``gfit den'' variant over the 24 sets of the Galaxy challenge. This variant has a preliminary denoising step using the DWT-Wiener algorithm. The red lines denote the estimated shear power spectra while the green lines represents the true shear power. The blue lines indicate the spectra after application of pixel-level denoising (not DWT-Wiener denoising). The code run to plot these power spectra is identical to that used in the \citep{G10results} paper.}
\label{fig:gfit den power spectra}
\end{center}
\end{figure*}

Figs~\ref{fig:gfit power spectra} and \ref{fig:gfit den power spectra} respectively show the shear power spectra of the non-denoised and denoised variants of \emph{gfit} submitted in the GREAT10 galaxy Challenge. 

\end{appendix}

\end{document}